\documentclass[12pt]{article}

\usepackage[dvips]{graphicx}
\usepackage{psfrag}
\usepackage{amssymb}

\newcommand{\eq}[1]{\begin{equation} #1 \end{equation}}
\newcommand{\eqn}[1]{\begin{displaymath} #1 \end{displaymath}}
\newcommand{\eqa}[1]{\begin{eqnarray} #1 \end{eqnarray}}

\newcommand{\av}[1]{\langle #1 \rangle}
\newcommand{\sss}{\scriptscriptstyle}

\newcommand{\heff}{\mathcal{H}_{\rm eff}}
\newcommand{\Aeff}{\mathcal{A}_{\rm eff}}
\newcommand{\ANLO}{\mathcal{A}_{\scriptscriptstyle \rm NLO}}
\newcommand{\op}{\mathcal{O}}
\newcommand{\nn}{\nonumber}
\newcommand{\Li}{{\rm Li_2}}

\begin{document}


$\ $
\vspace{1.5cm}
\begin{center}
\Large\bf 
Exact NLO strong interaction corrections to the\\ $\Delta F=2$ effective Hamiltonian in the MSSM
\end{center}

\vspace{3mm}
\begin{center}
{\sc Javier Virto
}
\end{center}

\begin{center}
{\em 
INFN, Sezione di Roma, I-00185 Rome, Italy
}
\end{center}

\vspace{1mm}
\begin{abstract}\noindent

We compute the order $\alpha_s^3$ (next-to-leading) corrections to the Wilson coefficients of the $\Delta F=2$ effective Hamiltonian in the Minimal Supersymmetric Standard Model with completely arbitrary soft terms. These results are relevant for phenomenological studies of neutral meson mixings in the presence of large squark mass splittings, in particular in hierarchical scenarios of supersymmetry. These corrections achieve a considerable reduction of the uncertainty related to the matching scale ambiguity, and are therefore numerically relevant. We briefly analyze the effect of certain mass splittings on the size of the NLO corrections, and compare it with the case of almost degenerate squarks in the mass insertion approximation.

\end{abstract}


\section{Introduction}

There is an almost absolute consensus among the particle physics community that unknown physics must exist at energies of around the TeV. Indeed, the minimal Higgs mechanism, so far completely untested, is quite unlikely to be the true mechanism for electroweak symmetry breaking; and even if it were, an additional explanation should be provided for the stabilization of the electroweak scale. Another indication of new physics around these scales is given by, for example, the dark matter problem. Low-energy supersymmetry is a leading candidate to fill that gap since it provides natural and simple explanations to those conundrums.

However, over the last decade, two other things have become clear. First, that the structure of flavor symmetry breaking of the Standard Model is a very peculiar one, and second, that, quite unexpectedly, flavor physics experiments confirm it to great accuracy. Indeed, any new physics model one can think of leads naturally to violent flavor breaking, and the experimental limits on flavor-changing neutral currents (FCNC's) are among the most stringent constraints that any model must satisfy.

The Minimal Supersymmetric Standard Model (MSSM) is no exception: an enormous set of new flavor violating parameters arise from its supersymmetry-breaking sector, which in the quark sector can be understood as a general misalignment between quark and squark mass matrices. More specifically, a simultaneous flavor rotation of quark and squark fields diagonalizing the Yukawa matrices leads to the so called super-CKM basis, in which no tree-level FCNC couplings are present. But in this basis the squark mass matrix $M^2$ is not generally diagonal, and an extra rotation of the squark fields is necessary to diagonalize this matrix, introducing strong interaction flavor violating couplings. Therefore, SUSY contributions to flavor violating observables mediated by strong interactions will easily compete with the SM contributions, which are driven by weak interactions, and well tested experimentally.  Flavor constraints on these flavor violating parameters have been studied extensively, and the conclusion to be taken from those studies is, basically, that it is quite difficult to reconcile the relatively low SUSY masses required by naturalness with a generic flavor structure of the soft breaking terms.

A solution to this problem is to assume that the structure of flavor violation of the new physics respects the hypothesis of Minimal Flavor Violation (MFV), by virtue of which the Yukawa couplings are the only source of flavor symmetry breaking. However, this is quite a pessimistic scenario for new physics searches in the quark sector, and it is useful for many purposes to go beyond it. Indeed, present indications of new physics in $B_{d,s}$ and $K$ processes \cite{arXiv:0803.4340,arXiv:0803.0659,arXiv:0805.3887} seem to require a departure from MFV.

In the MSSM, a suitable approach for phenomenological studies that takes into consideration all these issues, is to assume that squark masses are very nearly degenerate. This case is parametrized by a squark mass matrix of the form $(M^2)_{ij}=M^2_s(1+\delta)_{ij}$, where $M_s$ is an ``average'' squark mass, and $\delta$ is a matrix with entries much smaller than one. These small parameters, or \emph{mass insertions}, parametrize the departure from MFV, and constraints on their values derived from flavor observables provide valuable information for model building. They are usually treated as expansion parameters, which at leading order (linear and quadratic in $\delta$ for $\Delta F=1$ and $\Delta F=2$ processes respectively), define what is coventionally know as the \emph{Mass Insertion Approximation} (MIA) \cite{Hall:1985dx}. Strong bounds have been derived for these quantities from FCNC processes \cite{Ciuchini:1998ix,Becirevic:2001jj,Silvestrini:2007yf,Ciuchini:2007cw} and vacuum stability requirements \cite{hep-ph/9507294,hep-ph/9606237} as well as from charged-current processes \cite{arXiv:0810.1613}. 

However, it is also of interest to consider other scenarios with non-degenerate squark masses. For example, it has been argued that a ``hierarchical'' setup in which the first two generations of squarks are much heavier than the rest of the SUSY spectrum (lying near the electroweak scale), can satisfy naturalness criteria \cite{hep-ph/9607394}. In this framework, correlation patterns between  $\Delta F=1$ and $\Delta F=2$ observables can be quite different from the ones in the degenerate case. Indeed, as shown in Ref.~\cite{arXiv:0812.3610}, in these scenarios the bounds from $B\to X_s\gamma$ can be partially evaded,  allowing for a large $B_s$ mixing phase. This would be of great relevance if the experimental indication for such a large phase \cite{arXiv:0712.2397,arXiv:0802.2255,arXiv:0803.0659} is confirmed. 

In order to perform phenomenological studies of these scenarios, corresponding calculations of flavor violating processes have to be performed. Here we focus on $\Delta F=2$ processes, that is, $K-\bar K$, $D-\bar D$ and $B_{d,s}-\bar B_{d,s}$ mixing. These low energy observables are more conveniently computed in the framework of an effective theory in which heavy modes have been integrated out. The most general effective hamiltonian relevant for $\Delta F=2$ processes is given in Eq.~(\ref{Heff}), where the Wilson coefficients $C_i$ contain the information from heavy modes. Then the observables are expressed as functions of the Wilson coefficients and the matrix elements of the operators, and arise from the $\Delta F=2$ amplitude in the effective theory:
\eq{\Aeff=\sum_{i}C_i\,\av{\op_i}\,=\sum_{ij}C_i\,\left( \delta_{ij}+\frac{\alpha_s}{4\pi} r_{ij}+\op(\alpha_s^2)\right)\,\av{\op_j}^{(0)}\ ,
\label{Aeff}}
where we have written the matrix elements of the operators in terms of tree level matrix elements $\av{\op}^{(0)}$. The matrix elements must be computed using some non-perturbative approach, for example in the lattice. The SUSY contributions are encoded inside the Wilson coefficients, which are evaluated by matching the full theory (MSSM) onto the effective theory at some matching scale $\mu$. According to the renormalization group (RG) prescription for the resummation of large logarithms, the matching scale $\mu$ must be close to the SUSY scale, and the scale at which matrix elements are computed must be close to a relevant mass scale in the effective theory (for example $m_b$ in the case of $B-\bar B$ mixing), and the Wilson coefficients at the matching scale must be used as initial conditions for the RG evolution that provides the Wilson coefficients at the low scale. This evolution is governed in particular by the anomalous dimensions of the operators.

Leading order (LO) strong interaction matching conditions in the MSSM have been computed in Refs.~\cite{Gerard:1984bg,hep-ph/9604387,Hagelin:1992tc}, and arise from the squark-gluino box diagrams shown in Appendix~\ref{appLO}. The corresponding next-to-leading order (NLO) corrections arise from the two loop diagrams shown in Appendices~\ref{appIR} - \ref{appFIN}, and have been computed in Ref.~\cite{hep-ph/0606197} within the Mass Insertion Approximation. The anomalous dimension matrix for the complete set of operators (Eq.~(\ref{ops})) has been computed at NLO in QCD in Refs.~\cite{hep-ph/9711402,hep-ph/0005183}.

The purpose of this paper is to present the computation of the NLO matching conditions beyond the MIA, in the presence of arbitrary soft terms, and in particular for arbitrary squark mass splittings. The motivation for a NLO determination of the matching conditions is three-fold. First, LO matching conditions are both scale- and scheme-independent, so in order to get scheme independent results and NLO scale invariance it is necessary to go beyond the leading order. Second, the leading order corrections are proportional to $\alpha_s^2$. Since at LO neither the renormalization scale or the scheme can be specified for the strong coupling, LO results show a particularly high uncertainty related to the scheme and scale ambiguities. Third, this uncertainty is particularly severe due to the large anomalous dimensions of the $\Delta F=2$ operators involved. These uncertainties are largely cured by the NLO corrections, from about 10-15\% to a few percent, as shown explicitly in Ref.~\cite{hep-ph/0606197}. 

The $\Delta F=2$ amplitude in the MSSM up to NLO can be written as
\eq{\mathcal{A}_{\scriptscriptstyle\rm MSSM}=\sum_i \alpha_s^2 \left( F_i^{(0)}+\frac{\alpha_s}{4\pi}F_i^{(1)} +\op(\alpha_s^2)\right)\av{\op_i}^{(0)}\ ,
\label{Afull}}
where $F_i^{(0)}$ and $F_i^{(1)}$ are the LO and NLO contributions, and we have factored out a common $\alpha_s^2$. The matching of the full theory onto the effective theory is performed by imposing that the effective and MSSM amplitudes are equal at and below a matching scale $\mu$. An order by order identification of Eqs.~(\ref{Aeff}) and (\ref{Afull}) leads to the following formula for the Wilson coefficients:
\eq{
C_i=\alpha_s^2F_i^{(0)}+\frac{\alpha_s^3}{4\pi}F_i^{(1)}-\frac{\alpha_s^3}{4\pi}\sum_j F_j^{(0)}r_{ji}+\op(\alpha_s^4)\ .
\label{WC}}
Thus, the NLO matching calculation requires the computation of the matrix $r$ and the functions $F_i$. The matrix $r$ is obtained from the renormalization of the operators of the effective theory, and its computation is described in Section~\ref{sec:Heff}. This matrix has been computed before in several renormalization schemes (see Refs.~\cite{hep-ph/9711402,hep-ph/0005183,hep-ph/0606197}), and we agree with their results. The functions $F_i$ are obtained computing the one- and two-loop diagrams in the MSSM (see Appendix~\ref{app1}). The details of this computation are described in Section~\ref{sec:Det}, and the functions $F_i^{(1)}$ obtained here are the main new results of this paper.

Both calculations, the effective and the full theory amplitudes, are carried out in the NDR scheme (dimensional regularization with anticommuting $\gamma_5$), with modified minimal subtraction ($\rm \overline{MS}$) of ultraviolet divergencies. Also, we choose massless external quarks with zero external momenta. This choice of external states introduces infrared (IR) divergencies from diagrams in which a gluon connects two external legs. We regularize these divergencies with an unphysical gluon mass $\lambda$. While both $r$ and $F_i^{(1)}$ depend on the external states (and are therefore IR divergent), this dependence cancels in the Wilson coefficients, as expected. 

After describing the computation of the relevant amplitudes, we discuss briefly in Section~\ref{sec:ren} some issues related to the renormalization of ultraviolet divergencies and the renormalization scale dependence. In Section~\ref{sec:checks} we mention some of the checks that can be done to ensure the correctness of the results. In Section~\ref{sec:MIA} we show how reduce the exact results to the MIA and to the MIA with non-degenerate squarks, which allows to compare our results with those in Ref.~\cite{hep-ph/0606197}. Finally, some results are presented in Section~\ref{sec:res}.

Before getting down to brass tacks, we would like to set some notation and specify some definitions related with rotation and mass matrices beyond leading order. At tree level, the super-CKM basis is defined by doing a joint rotation in flavor space of quark and squark fields such as to diagonalize the tree level Yukawa matrices. The resulting squark mass matrix is not diagonal and defines the tree level mass insertions. In this basis there are no tree level FCNC's. This matrix is diagonalized by an additional rotation of the squark fields,
\eqa{
\tilde d_{i,L}^I=\Gamma_{D_L}^{j i*}\tilde d_{j}\ ,&\quad& \tilde d_{i,R}^I=\Gamma_{D_R}^{j i*}\tilde d_{j}\nn\\
\tilde u_{i,L}^I=\Gamma_{U_L}^{j i*}\tilde u_{j}\ ,&\quad& \tilde u_{i,R}^I=\Gamma_{U_R}^{j i*}\tilde u_{j}
}
where $(\tilde q^I_L,\tilde q^I_R)$ denote the squark fields in the (tree level) super-CKM basis, and $\tilde q$ is the mass eigenbasis. The rotation matrices $\Gamma_L$ and $\Gamma_R$ are $3\times 6$ matrices, and the indices $U,D$ will be omitted hereon, which raises no confusion. In the mass eigenbasis, tree level FCNC's appear; for example a flavor changing $\tilde q_i$-$q_j$-$\tilde g$ vertex is generated with the following Feynman rule:
\eq{-i g_s\sqrt{2} T^a(\Gamma_L^{ij}P_L-\Gamma_R^{ij}P_R)\ ,}
where $T^a$ are the color matrices and $P_{L,R}$ are the chiral projectors.

At NLO, a subtlety arises because a squark-gluino loop generates a finite flavor-changing self energy for the quark fields (see Fig.~\ref{FCSE}). At this point, one must specify what is meant by the super-CKM basis at NLO, since the definitions for the mass insertions depend on that choice. We believe that the most natural definition for the super-CKM basis is the one for which quark fields do not mix at one loop, and tree-level FCNC's are absent. In this case the rotation matrices, the mass insertions and the CKM matrix differ from the tree level ones. This criterium does not coincide with that in Ref.~\cite{arXiv:0810.1613}, but it has no effect when comparing our results with Ref.~\cite{hep-ph/0606197}, since this is not an issue in the degenerate case. A comment in favor of the criterium adopted in Ref.~\cite{arXiv:0810.1613} is that mass insertions are directly related with SUSY-breaking parameters in the lagrangian. A further discussion on this issue is provided in Section~\ref{sec:ren}, and the formulae necessary to switch from one criterium to the other is provided in Appendix \ref{app3}.

\section{Effective Hamiltonian for $\Delta F=2$ processes at NLO}
\label{sec:Heff}

The most general effective Hamiltonian for  $\Delta F=2$ processes up to operators of dimension six can be written as
\eq{\heff^{\Delta F=2}=\sum_{i=1}^5 C_i\, \op_i+\sum_{i=1}^3\tilde{C}_i\,\tilde{\op}_i
\label{Heff}}
where $C_i$ are the Wilson coefficients and $\op_i$ are the dimension six $\Delta F=2$ operators. In four dimensions there are eight independent operators of this type. Here we choose the following basis:
\eqa{
\op_1& = & \bar s_\alpha \gamma_\mu P_Lb_\alpha\ \bar s_\beta \gamma^\mu P_Lb_\beta \nn\\
\op_2& = & \bar s_\alpha P_Lb_\alpha\ \bar s_\beta P_Lb_\beta \nn\\
\op_3& = & \bar s_\alpha P_Lb_\beta\ \bar s_\beta P_Lb_\alpha \nn\\
\op_4& = & \bar s_\alpha P_Lb_\alpha\ \bar s_\beta P_Rb_\beta \nn\\
\op_5& = & \bar s_\alpha P_Lb_\beta\ \bar s_\beta P_Rb_\alpha \label{ops}
}
where $P_{L,R}=(1\mp\gamma_5)/2$ are the usual chiral projectors. The operators $\tilde\op_{1,2,3}$ are obtained from $\op_{1,2,3}$ by exchanging $L\leftrightarrow R$. To simplify the notation throughout the paper we focus on the case of $B_s-\bar B_s$ mixing; for the cases of $K$, $D$ and $B_d$ mixing one should make obvious replacements of the quark fields.

In order to define the effective Hamiltonian beyond leading order, one must specify a renormalization scheme. Here we choose to regularize ultraviolet divergencies in dimensional regularization, where $d=4$ identities between operators do not hold. This means that one must complete the effective Hamiltonian with a set of evanescent operators that vanish in 4 dimensions but can give finite contributions beyond leading order if they are multiplied by a divergence. The choice of a set of evanescent operators is not unique, and different sets lead to different subtractions, so specifying this set is necessary to fix the renormalization scheme. Here we choose the set of evanescent operators given in Ref.~\cite{hep-ph/0005183}. By requiring that the matrix elements of evanescent operators vanish in four dimensions one can omit these operators altogether from the effective Hamiltonian once the renormalization has been performed.

\begin{figure}
\begin{center}
\includegraphics[height=2.5cm]{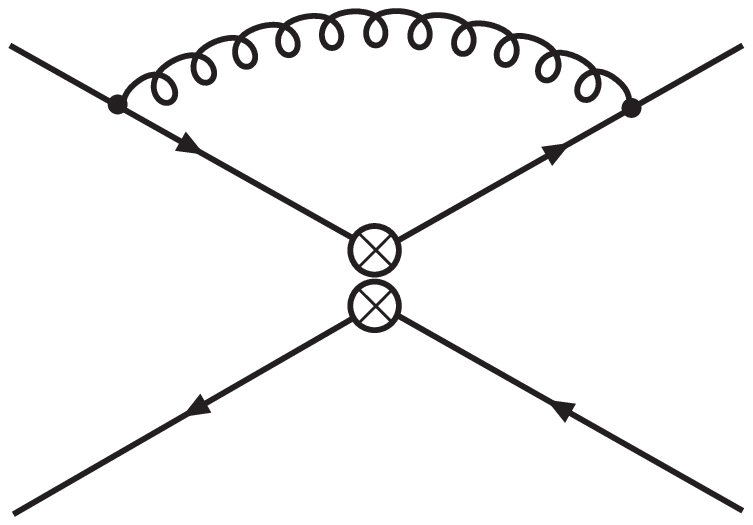}\hspace{1cm}
\includegraphics[height=2.3cm]{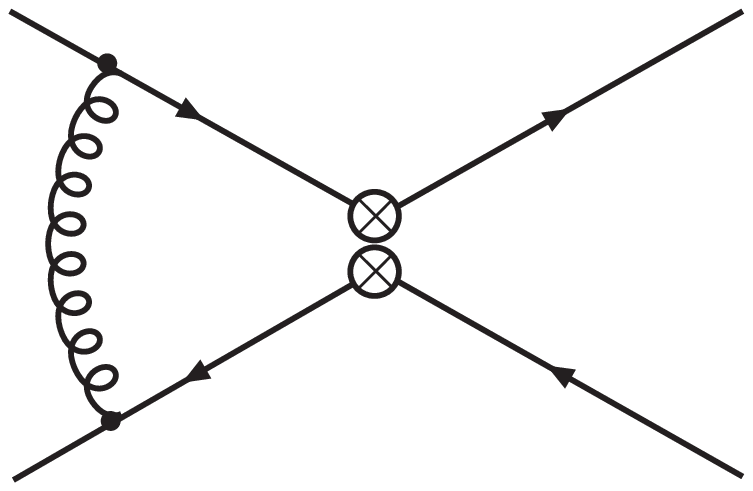}\hspace{1cm}
\includegraphics[height=2.5cm]{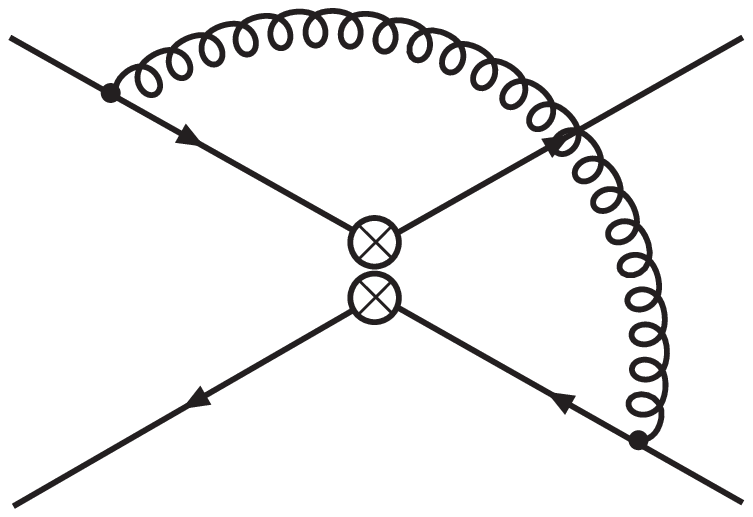}
\end{center}
\caption{\small One loop diagrams contributing to the matrix elements of the four-fermion operators in the effective theory.}
\label{DiagsEff}
\end{figure}

The calculation of the NLO effective Hamiltonian amounts basically to the NLO renormalization of the operators. The matrix elements of the bare operators and of the renormalized operators are related through the renormalization constants:
\eq{\av{\op_i}^{\rm bare}=\sum_j Z_{ij} Z_s Z_b\,\av{\op_j}^{\rm ren}=\sum_j Z_{ij}'\,\av{\op_j}^{\rm ren}}
where $Z_{s,b}$ are the quark wave function renormalization factors and $Z_{ij}$ is the renormalization matrix necessary to renormalize properly the operators in the effective theory. The effective theory amplitude can then be written as
\eq{\Aeff=\sum_i C_i\,\av{\op_i}^{\rm ren}+\sum_{i,j} C_i\,\delta Z_{ij}'\av{\op_i}^{\rm ren} }
where we have written $Z_{ij}'=\delta_{ij}+\delta Z_{ij}'$ and the second term contains the counterterms. In order to obtain the NLO renormalized operators $\av{\op_i}^{\rm ren}$ in terms of tree level matrix elements $\av{\op_i}^{(0)}$, one must compute the one loop gluonic corrections such as those shown in Fig. \ref{DiagsEff}. For UV divergencies we use naive dimensional regularization with modified minimal substraction ($\overline{\rm MS}$-NDR). Moreover, we choose zero external momenta and set the quark masses to zero. This simplifies the computation but introduces IR divergencies from the soft gluon region. We regularize this divergencies using a gluon mass $\lambda$. The same IR divergencies should appear in the full theory, and cancel in the matching, providing a non-trivial check of the calculation. 

All the one loop diagrams contributing to the NLO renormalized operators are proportional to the same loop integral, and the amplitude can be written as
\eq{\Aeff=\sum_{ij} C_i\,\left[\, \delta_{ij}-  \frac{\alpha_s}{16\pi}\bigg( \frac{1}{\hat\epsilon}+\frac{3}{2}-2 \log (\lambda/\mu)\bigg)\,A_{ij}+\delta Z_{ij}' \,\right]\ \av{\op_j}^{(0)}\ .}
The computation then provides the coefficients $A_{ij}$, which must be extracted up to $\op(\epsilon)$: $A_{ij}=A_{ij}^0+\hat\epsilon A_{ij}^\epsilon$. The sum over $j$ runs over both physical and evanescent operators, the later giving finite contributions to $A_{ij}^\epsilon$.

The counterterms can now be fixed according to the $\overline{\rm MS}$ scheme: $\delta Z_{ij}' =(\alpha_s/16\pi\hat\epsilon) A_{ij}^0$, which provides the leading order anomalous dimension matrix of the operators (\ref{ops}) in QCD:
\eq{\gamma\equiv Z^{-1}\frac{d Z}{d\log\mu}=\frac{\alpha_s}{4\pi}\gamma^{(0)}=\frac{\alpha_s}{4\pi}\,\bigg[ 4C_F-\frac{1}{2}A^0 \bigg]}
where we have used the well known QCD quark wave function renormalization factor, $Z_q=1-(\alpha_s/4\pi\hat\epsilon)\,C_F$. We obtain
\eq{\gamma^{(0)}=
 \left(
\begin{array}{ccccc}
4 &        0                      &               0                &                0             &               0 \\
0 &        -28/3                      &          4/3                       &                0              &          0        \\
0 &          16/3                      &               32/3               &           0                      &               0 \\
0 &        0                      &               0               &               -16              &               0 \\
0 &        0                      &            0                    &                -6              &           2       \\
\end{array}
\right)
}
in agreement with, for example, refs. \cite{hep-ph/9707225,hep-ph/0606197}. The missing 3$\times$3 block corresponding to the operators $\tilde \op_{1,2,3}$ has been omitted: they do not mix with the other operators and their anomalous dimensions are the same as for $\op_{1,2,3}$.

Finally, the NLO amplitude in the effective theory is given by
\eqa{
\Aeff&=&\sum_{ij}C_i\,\left( \delta_{ij}+\frac{\alpha_s}{4\pi} r_{ij}\right)\,\av{\op_j}^{(0)}\label{Aeff2}\\
r_{ij}&=&\frac{1}{4}\left[ \bigg( \frac{3}{2}-2 \log (\lambda/\mu) \bigg)A_{ij}^0+A_{ij}^\epsilon \right]\ .
}
For the NLO matrix $r$ we obtain
\eqn{
r_{ij}=
 \left(
\begin{array}{ccccc}
-4/3 &        0                      &               0                &                0             &               0 \\
0 &        -44/3                      &          4/3                       &                0              &          0        \\
0 &          16/3                      &               16/3               &           0                      &               0 \\
0 &        0                      &               0               &               -64/3              &               0 \\
0 &        0                      &            0                    &                -6              &           -10/3       \\
\end{array}
\right) \log(\lambda/\mu)+
\left(
\begin{array}{ccccc}
-5 &        0                      &               0                &                0             &               0 \\
0 &        1/3                      &          -1                       &                0              &          0        \\
0 &          -15/2                      &               -25/6               &           0                      &               0 \\
0 &        0                      &               0               &               19/3              &               -3 \\
0 &        0                      &            0                    &                -1/2              &           -7/6       \\
\end{array} \right)
}
The first term is the IR divergent piece that must cancel in the matching procedure. The second term is the NLO contribution, and it is scheme dependent: it is valid only in the $\overline{\rm MS}$-NDR scheme. However it can be easily translated to dimensional reduction (DRED) and Regularization-Independent (RI) schemes by using the formulae in Refs.  \cite{hep-ph/0606197,hep-ph/9711402}. In fact, as pointed out in Ref.~\cite{hep-ph/9711402}, the matrix $r$ can be thought of as \emph{defining} the renormalization scheme, and Eq.~(\ref{Aeff2}) as a definition the renormalized operators. The same scheme must be used in the full theory calculation and in the evaluation of the matrix elements in order to obtain scheme-independent results for physical amplitudes.

\section{Details of the calculation in the MSSM}
\label{sec:Det}

The NLO $\Delta F=2$ amplitude in the MSSM is obtained by computing the one- and two-loop Feyman diagrams listed in Appendix \ref{app1}. The one-loop contributions are the box diagrams shown in appendix \ref{appLO}; the computation of these graphs is standard and will not be reviewed here. Renormalization of the two-loop diagrams requires the inclusion of these boxes with vertex counterterms, as well as pentagons containing the self-energy counterterms. These one loop integrals must then be computed up to and including terms of $\op(\epsilon)$, which provide the finite scale- and scheme-dependent contributions to the NLO amplitude. We will come back to this in section \ref{sec:ren}.

The computation of the two-loop diagrams is done in four steps:

\begin{enumerate}
\item Partial fractioning of the denominators and tensor reduction to reduce the Feynman integrals to a set of scalar integrals with three scalar propagators.
\item Decomposition of the scalar integrals down to a set of one- and two-loop master integrals. These master integrals are known functions of masses and contain divergencies up to $\op(1/\epsilon^2)$.
\item Cancellation of UV divergencies.
\item Fierz rearrangement and spinor transpositions in order to express the spinor structures in terms of tree level matrix elements of the operators.
\end{enumerate}

Let us describe in some detail each of these steps.\\

{\bf 1.} Since we are choosing massless quarks with zero external momenta, there are only two independent momenta appearing in the Feynman integrals. One can then decompose the denominators such that each diagram can be expressed as a sum of terms of the following type
\eqn{f(m{\rm 's}) \ (\bar s_\beta\, \Gamma^{\mu_1,\nu_1,\dots}_{\alpha,\beta,\dots}\, b_\alpha)\ (\bar s_\gamma\, \bar\Gamma^{\mu_2,\nu_2,\dots}_{\alpha,\beta,\dots}\, b_\delta)\ 
\int \frac{d^Dq_1}{(2\pi)^D} \frac{d^Dq_2}{(2\pi)^D} \frac{q_1^{\mu_1} q_1^{\mu_2} \cdots q_2^{\nu_1} q_2^{\nu_2} \cdots}{(q_1^2-m_1^2)^{n_1} (q_2^2-m_2^2)^{n_2} (\Delta q^2-m_3^2)^{n_3}}
}
where the $f$'s are some functions that depend in general on all the masses appearing in the Feynman diagram (including the fictitious gluon mass),  and $\Delta q\equiv q_1-q_2$. Also, $\Gamma$ and $\bar \Gamma$ represent Dirac and color structures.  The spinors $s$ and $b$ might be $u$ or $v$ spinors and they might appear transposed and in different order, according to the chosen reference order adopted to keep track of the relative signs of interfering Feynman graphs \cite{Denner:1992vza}.

The tensor integrals are momentum-independent (again, because external momenta are zero), so they can be expressed in terms of scalar integrals multiplied by metric tensors. For example,
\eqn{
\int \frac{d^Dq_1}{(2\pi)^D} \frac{d^Dq_2}{(2\pi)^D} \frac{q_1^{\mu} q_2^{\nu}}{(\cdots)} =
\frac{g^{\mu\nu}}{D}\int \frac{d^Dq_1}{(2\pi)^D} \frac{d^Dq_2}{(2\pi)^D} \frac{q_1\cdot q_2}{(\cdots)}
}
In this way we can express each diagram as a sum of terms of the form
\eqn{f(m{\rm 's}) \ (\bar s_\beta\, \Gamma^{\mu,\nu,\dots}_{\alpha,\beta,\dots}\, b_\alpha)\ (\bar s_\gamma\, \bar\Gamma^{\mu,\nu,\dots}_{\alpha,\beta,\dots}\, b_\delta)\ 
\int \frac{d^Dq_1}{(2\pi)^D} \frac{d^Dq_2}{(2\pi)^D} \frac{(q_1^2)^a\, (q_2^2)^b\, (q_1\cdot q_2)^c}{(q_1^2-m_1^2)^{n_1} (q_2^2-m_2^2)^{n_2} (\Delta q^2-m_3^2)^{n_3}}
}
At this point, care must be taken when manipulating the Dirac structures $\Gamma$ and $\bar \Gamma$ after contraction with the metric tensors. In particular, in $D=4-2\epsilon$ dimensions, structures of the type $(\gamma_\mu\gamma_\nu\gamma_\lambda\cdots)\otimes(\gamma^\mu\gamma^\nu\gamma^\lambda\cdots)$ cannot be reduced like in $D=4$, and evanescent structures (of order $\op (\epsilon)$) must be introduced. According to the NDR prescription here adopted, however, one can freely anti-commute the $\gamma_5$ and use the usual anti-commutation relations for $\gamma$ matrices.\\

{\bf 2.} The resulting scalar integrals can be further reduced to a set of one- and two-loop master integrals by the method of recurrence relations  \cite{hep-ph/9703319}. This reduction can be performed automatically using the Mathematica program TARCER  \cite{hep-ph/9801383}. In this way the scalar integrals can be expressed in terms of various one-loop tadpole integrals and a single two-loop master integral,
\eq{\mathcal{I}(m_1,m_2,m_3)\equiv \int   \frac{d^Dq_1\,d^Dq_2}{(q_1^2-m_1^2) (q_2^2-m_2^2) (\Delta q^2-m_3^2)}\ .}
The result for this master integral with arbitrary masses is given in ref.  \cite{Davydychev:1992mt}.\\

{\bf 3.} Once all the loop integrations have been performed, the resulting expression for the Feynman diagram consists of a sum of terms of the form
\eq{
\left(\frac{ A(m_{\tilde g},\tilde m)}{\epsilon^2} + \frac{ B(m_{\tilde g},\tilde m)}{\epsilon} + C(m_{\tilde g},\tilde m)\right)\ 
(\bar s_\beta\, \Gamma^{\mu,\nu,\dots}_{\alpha,\beta,\dots}\, b_\alpha)\ (\bar s_\gamma\, \bar\Gamma^{\mu,\nu,\dots}_{\alpha,\beta,\dots}\, b_\delta)
}
where $A$, $B$ and $C$ are some functions of gluino and squark masses. Since in this case one-loop corrections are finite, no $1/\epsilon^2$ divergencies can appear, so all $1/\epsilon^2$ terms should (and do) cancel directly. We also get automatic cancellation of $1/\epsilon$ terms for the diagrams shown in appendix \ref{appFIN}, as it should be. The rest of the diagrams (those shown in appendix \ref{appUV}) contain $1/\epsilon$ divergencies that cancel against one-loop diagrams with vertex and self-energy counterterms. We will discuss the details of the renormalization in section \ref{sec:ren}.\\

{\bf 4.}  After all the divergencies have been removed, 4D Fierz identities can be used and transposition of spinors can be performed to put all spinor and Dirac structures in suitable form. These structures must then appear in the precise combinations that constitute the tree level matrix elements of the physical operators, as for example,
\eqn{
2\,(\bar u_s^\alpha\gamma_\mu P_L v_b^\alpha)\,(\bar v_s^\beta\gamma_\mu P_L u_b^\beta) + 2\,(\bar u_s^\alpha\gamma_\mu P_L v_b^\beta)\,(\bar v_s^\beta\gamma_\mu P_L u_b^\alpha)
\longrightarrow  \av{\op_1}^{(0)}\ .
}
This, however, does not occur for individual diagrams, but only for certain groups of diagrams (and of course for the amplitude as a whole). Therefore this step provides an interesting check of the calculation, in particular of the relative sign between the different diagrams. The NLO quantities $F_i^{(1)}$ in eq. (\ref{Afull}) are then obtained by summing all the contributions.

\section{Renormalization}
\label{sec:ren}

In order to reduce the ultraviolet divergencies arising from the two-loop graphs, appropriate counterterms must be introduced. In particular, the squark-quark-gluino vertex, as well as the quark, squark and gluino wave functions and masses have to be renormalized. For reference, the relevant renormalization factors in the $\overline{\rm MS}$-NDR scheme up to $\op(\alpha_s)$ are
\eqn{
Z_{\tilde g} =  1+\frac{\alpha_s}{4\pi}\frac{1}{\hat \epsilon}\, (N_c+n_f)\ ,\quad
Z_{m_{\tilde g}} = 1-\frac{\alpha_s}{4\pi}\frac{1}{\hat \epsilon}\,  4 N_c\ ,
}
\eq{
Z_{\hat g_s} = 1-\frac{\alpha_s}{4\pi}\frac{1}{\hat \epsilon}\, (2 N_c+C_F)\ ,\quad
Z_{\tilde q} =  1 + \op(\alpha_s^2)\ ,\quad
Z_q =  1 -\frac{\alpha_s}{4\pi}\frac{1}{\hat \epsilon}\,2C_F\ ,
\label{Zfacts}}
which are defined in the usual way. Also, a non-diagonal squark-mass counterterm is necessary, because at one loop divergent flavor-changing squark propagators are generated. There are two ways of dealing with this issue: 1) Renormalize the squark rotation matrices so that non-diagonal squark masses are zero at one loop. In this case all the finite pieces of the one loop corrections cancel with the renormalized non-diagonal mass insertions, which are not zero anymore but of order $\op(\alpha_s)$. 2) Renormalize the non-diagonal squark-mass parameters minimally, and include the finite pieces of the loop corrections to the flavor-changing squark propagator. In this case the squark rotation matrices are defined such that renormalized mass insertions are zero at the matching scale.

We choose the second option, and renormalize the squark mass parameters according to $m_{ij}^{2\ \rm bare}=m_{ij}^2+\delta_{m_{ij}}$, with
\eqa{
&&\delta_{m_{ij}}  =  -\frac{\alpha_s}{4\pi}\frac{1}{\hat \epsilon}\ 2 C_F \bigg[ (\tilde m_i^2+2m_{\tilde g}^2) \delta_{ij} - 
\sum_{k,q,q'}\tilde m_k^2\,(\Gamma_L^{iq*} \Gamma_L^{kq} \Gamma_L^{kq'*} \Gamma_L^{jq'}+\Gamma_R^{iq*} \Gamma_R^{kq} \Gamma_R^{kq'*} \Gamma_R^{jq'}) \bigg]\ ,\nn\\
&&m_{ij}(\mu_{\scriptscriptstyle \rm Matching})  =  0\ .\hspace{10.5cm}
\label{renmij}
}
A consequence of all this is that, even though we are working in the basis of diagonal squark masses, non-diagonal mass parameters do run with the scale and contribute to the renormalization group equation.

There is an additional issue related with the fact that NDR is a regularization scheme that breaks supersymmetry. In particular, the coupling $g_s$ appearing in the quark-gluon-gluon vertex and the coupling $\hat g_s$ that appears in the quark-squark-gluino vertex, receive different radiative corrections in the NDR scheme. In the effective theory only $g_s$ appears, so when doing the matching it is convenient to have the MSSM amplitude expressed solely in terms of this coupling. Being this is an $\op(\alpha_s)$ effect, one can set $g_s=\hat g_s$ in the two-loop amplitude, but it gives a finite contribution of $\op(\alpha_s^2)$ from the LO amplitude. At the end, this effect is corrected for by performing in the one loop amplitude the following replacement (see  \cite{hep-ph/9308222, hep-ph/0606197}),
\eq{\hat g_s^{\rm NDR}=g_s^{\rm NDR} \left( 1+\frac{\alpha_s}{4\pi} \frac{4}{3} \right)\ .}

Finally, one must address the wave-function renormalization of the external states. There are two types of contributions: gluon corrections and squark-gluino corrections, both contributing a factor of $-C_F$ to $Z_q$ (see eq.~(\ref{Zfacts})). In addition, squark-gluino corrections contribute finite pieces of two types:
\begin{enumerate}
\item Flavor diagonal corrections:

These give a finite contribution to the on-shell quark wave-function renormalization constant,
\eq{
\delta Z_{q_{L,R}}=\frac{\alpha_s}{4\pi}\,C_F\, \Bigg[  \log (m_{\tilde g}^2/\mu^2)-\sum_{k} f(\tilde m_k/m_{\tilde g})\, \Gamma_{L,R}^{kq*}\, \Gamma_{L,R}^{kq}\Bigg]\ ,
}
with $f(x)=(x^2-4x+3-2x(x-2)\log x)/2 (x-1)^2$.

\item Flavor changing corrections: 

The presence of quark flavor changing squark-gluino loops (see Fig.~\ref{FCSE}) can be handled in two different ways, in relation to two different definitions of the super-CKM basis: 

1) Tree-level definition of the super-CKM basis: This is the usual definition, in which the tree level Yukawa matrices are diagonal. At one loop, quarks of different flavors mix through the radiative corrections in Fig.~\ref{FCSE}, and one must include these corrections in the external legs.

2)``On-shell'' definition of the super-CKM basis: In this case the quark superfields are (finitely) renormalized with matrix-valued renormalization factors, that induce a perturbative rotation in flavor space. These counterterms are defined at each order in perturbation theory such as to render the quark self-energies diagonal at the given order, effectively canceling the flavor-changing self-energies of Fig.~\ref{FCSE}. 

\begin{figure}
\begin{center}
\psfrag{q1}{$q$} \psfrag{q2}{$q'$}\psfrag{s}{$\tilde q$} \psfrag{g}{$\tilde g$}
\includegraphics[height=3cm,width=6cm]{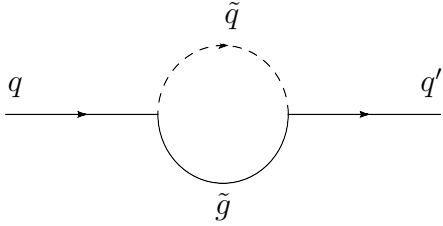}
\end{center}
\caption{Flavor changing quark self energies mediated by a squark-gluino loop.}
\label{FCSE}
\end{figure}

As mentioned in the introduction, the mass insertions (and the squark rotation matrices) that arise in each scheme represent different quantities.  Here we assume an on-shell definition of the super-CKM basis, and omit the flavor-changing external-leg corrections in our computations. For completeness, we provide in Appendix \ref{app3} the explicit relationship between squark rotation matrices and mass insertions in both definitions. The full results for the NLO Wilson coefficients consistent with the usual tree-level definition of the super-CKM basis can be obtained by adding an $\op(\alpha_s)$ correction to the rotation matrices $\Gamma_{L,R}^{ij}$ in the LO results, as explained in detail in Appendix \ref{app3}.

\end{enumerate}

Once all the renormalization factors have been defined, one can write down explicitly the renormalization group equation for the Wilson coefficients,
\eq{\left[  \frac{\partial}{\partial \log \mu^2}+\frac{d\,\alpha_s}{d \log \mu^2} \frac{\partial}{\partial \alpha_s}+\frac{d\, m_{\tilde g}^2}{d \log \mu^2} \frac{\partial}{\partial m_{\tilde g}^2}+
\sum_{i,j} \frac{d\,\tilde m_{ij}^2}{d \log \mu^2} \frac{\partial}{\partial \tilde m_{ij}^2}-\frac{1}{2}\gamma^T\,\right]\,\vec{C}(\mu)=0\ .
}
The renormalization group functions are given by
\eq{\frac{d\,\alpha_s}{d \log \mu^2}=-\frac{\alpha_s^2}{4\pi}\,(3 N_c-n_f)\ ,\quad \frac{d\, m_{\tilde g}^2}{d \log \mu^2}=\frac{\alpha_s^2}{4\pi}\,m_{\tilde g}^2\,(2n_f-6 N_c)\ ,
}
\eqn{
\frac{d\,\tilde m_{ij}^2}{d \log \mu^2} =  -\frac{\alpha_s}{4\pi}\ 2 C_F \bigg[ (\tilde m_i^2+2m_{\tilde g}^2) \delta_{ij} - 
\sum_{k,q,q'}\tilde m_k^2\,(\Gamma_L^{iq*} \Gamma_L^{kq} \Gamma_L^{kq'*} \Gamma_L^{jq'}+\Gamma_R^{iq*} \Gamma_R^{kq} \Gamma_R^{kq'*} \Gamma_R^{jq'}) \bigg]\ .
}
Therefore, with the notation of Eq.~(\ref{Afull}), the explicit form of the NLO RG-equation reads
\eq{
\frac{\partial}{\partial \log \mu^2} F_l^{(1)}=\hspace{12cm}
\label{RGE}
}
\eqn{2\,C_F\,[(\tilde m_i^2+2m_{\tilde g}^2) \delta_{ij} - 
\tilde m_k^2\,(\Gamma_L^{iq*} \Gamma_L^{kq} \Gamma_L^{kq'*} \Gamma_L^{jq'}+\Gamma_R^{iq*} \Gamma_R^{kq} \Gamma_R^{kq'*} \Gamma_R^{jq'})]\, \frac{\partial}{\partial \tilde m_{ij}^2} F_l^{(0)}+\frac{1}{2}\gamma_{il}^{(0)}F_i^{(0)}}
where appropriate sums over $i,j,k,q,q'$ are understood.

\section{Checks of the calculation}
\label{sec:checks}

There are several checks that can be made of the NLO calculation:

\begin{enumerate}

\item Cancellation of UV divergencies: All the UV divergencies arising from the two-loop graphs (specifically those in appendix \ref{appUV}) must be cancelled by one loop graphs with the insertions of the counterterms specified by the renormalization factors in eqs.~(\ref{Zfacts}) and (\ref{renmij}). 

\item Projection onto tree-level matrix elements: The NLO amplitude must be expressible in terms of tree-level matrix elements of the physical operators, that is, $\ANLO=F_i^{(1)} \av{\op_i}^{(0)}$. As mentioned before, this does not happen for individual two-loop graphs, and provides a check of interference between diagrams.

\item Cancellation of IR divergences: Both the amplitude in the effective theory and the amplitude in the MSSM depend on $\log \lambda$, where $\lambda$ is the gluon mass introduced to regularize the IR divergencies. This dependence must cancel completely in the matching when combining both amplitudes; that is, $F_i^{(1)}-F_k^{(0)} r_{ki}$ must be IR-finite. This is a non trivial check involving three completely independent pieces of the calculation: the matrix $r$ from the renormalization of the effective operators, $F_i^{(0)}$ from the one-loop graphs, and $F_i^{(1)}$ from the two-loop graphs.

\item NLO renormalization scale independence: The RGE (eq.~(\ref{RGE})) must be fullfilled. There is a close relationship between cancellation of UV divergencies and fulfillment of the RGE, but it is nevertheless a convenient check.

\end{enumerate}

We have verified that our results fulfill these requirements. Moreover, a final check consists in reducing the results to the degenerate MIA and comparing them with the results obtained in ref.~\cite{hep-ph/0606197}. In the next section we address the issue of how to reduce the exact results to the MIA and the NDMIA.

\section{Reduction to MIA and NDMIA}
\label{sec:MIA}

\subsection{Reduction to the Mass Insertion Approximation}
\label{sec:MIA1}

Any Wilson coefficient computed in this paper has the structure of a sum of terms of the following type,
\eq{
\begin{array}{l}
f_4(\tilde m_i^2,\tilde m_j^2)\cdot \Gamma_A^{is*}\,\Gamma_B^{ib}\cdot \Gamma_C^{js*}\,\Gamma_D^{jb}\\[2mm]
f_{6,1}(\tilde m_i^2,\tilde m_j^2,\tilde m_k^2)\cdot \Gamma_A^{is*}\,\Gamma_B^{ib}\cdot \Gamma_C^{js*}\,\Gamma_D^{jq}\cdot \Gamma_E^{kq*}\,\Gamma_F^{kb}\\[2mm]
f_{6,2}(\tilde m_i^2,\tilde m_j^2,\tilde m_k^2)\cdot \Gamma_A^{is*}\,\Gamma_B^{ib}\cdot \Gamma_C^{js*}\,\Gamma_D^{jb}\cdot \Gamma_E^{kq*}\,\Gamma_F^{kq}\\[2mm]
f_{8,1}(\tilde m_i^2,\tilde m_j^2,\tilde m_k^2,\tilde m_l^2)\cdot \Gamma_A^{iq*}\,\Gamma_B^{ib}\cdot \Gamma_C^{js*}\,\Gamma_D^{jq}\cdot \Gamma_E^{ks*}\,\Gamma_F^{kq'}\cdot \Gamma_G^{lq'*}\,\Gamma_H^{lb}\\[2mm]
f_{8,2}(\tilde m_i^2,\tilde m_j^2,\tilde m_k^2,\tilde m_l^2)\cdot \Gamma_A^{iq*}\,\Gamma_B^{ib}\cdot \Gamma_C^{js*}\,\Gamma_D^{jb}\cdot \Gamma_E^{ks*}\,\Gamma_F^{kq'}\cdot \Gamma_G^{lq'*}\,\Gamma_H^{lq}
\end{array}
\label{strWC}
}
where $A,B,C,\dots$ are either $L$ or $R$, and a sum is understood running over the indices $i,j,k,l=\tilde d_L, \tilde s_L, \tilde b_L, \tilde d_R, \tilde s_R, \tilde b_R$, and $q,q'=d,s,b$, (for $f_{6,2}$ also $q=u,c,t$ and
$k=\tilde u_L, \tilde c_L, \tilde t_L, \tilde u_R, \tilde c_R, \tilde t_R$).

These terms can be translated into functions of the entries of the squark mass matrix in the super-CKM basis $(M^2)$ using the following relations,
\eq{
\sum_i \Gamma_A^{iq*}\,\Gamma_B^{iq'}=\delta_{AB}\delta_{qq'}\quad ;\qquad \sum_i \tilde m_i^{2n}\,\Gamma_A^{iq*}\,\Gamma_B^{iq'}=[(M^2)^n]_{qq'}^{AB}\ .
\label{rotrel}
}
In the MIA, the diagonal elements in $M^2$ are assumed to be equal to an ``average'' squark mass $M_s^2$, and the off-diagonal elements (called mass insertions and denoted by $\Delta_{qq'}^{AB}$), are assumed to be much smaller than $M_s^2$. In this way, a power expansion on the \emph{dimensionless} mass insertions $\delta_{qq'}^{AB}\equiv \Delta_{qq'}^{AB}/M_s^2 \ll 1$ can be made, keeping in this case only the leading terms $\delta^2$. The mass eigenvalues are then $M_s^2(1+\op(\delta))$, so the functions in Eq.~(\ref{strWC}) can be Taylor-expanded around the average squark mass:
\eqa{
f(\tilde m_i^2,\tilde m_j^2,\dots)&=&f(M_s^2,M_s^2,\dots)+f^{(1,0,\dots)}(M_s^2,M_s^2,\dots)\cdot (\tilde m_i^2-M_s^2)+\cdots \nn \\
&&+\ f^{(1,1,0,\dots)}(M_s^2,M_s^2,\dots)\cdot (\tilde m_i^2-M_s^2) (\tilde m_j^2-M_s^2)+\cdots
}
Using the relations (\ref{rotrel}) it is easy to see that, for example,
\eqa{
f_{6,1}(\tilde m_i^2,\tilde m_j^2,\tilde m_k^2)\cdot \Gamma_L^{is*}\,\Gamma_R^{ib}\cdot \Gamma_L^{js*}\,\Gamma_L^{jq}\cdot \Gamma_R^{kq*}\,\Gamma_R^{kb}&=&\\[2mm]
f_{6,1}^{(1,0,1)}(M_s^2,M_s^2,M_s^2)\,\Delta_{sb}^{LR}\Delta_{sb}^{RR}
&+&f_{6,1}^{(1,1,0)}(M_s^2,M_s^2,M_s^2)\,\Delta_{sb}^{LR}\Delta_{sb}^{LL}\ , \nn
}
and similar for the other terms.

We have checked that our results, when reduced to the MIA, reproduce exactly the results in Ref.~\cite{hep-ph/0606197}.

\subsection{Reduction to the Non-degenerate MIA}
\label{sec:MIA2}

From the exact results obtained in this paper, one of the most straightforward generalizations of the MIA results that can be obtained is the case in which the diagonal elements of the squark mass matrix are non-degenerate. In this case, squark masses can be widely different while keeping mass insertions small. We call this the \emph{Non-degenerate Mass Insertion Approximation} (NDMIA), in which diagonal masses are non-degenerate but mass insertions are kept only up to second order.

In order to reduce the terms in Eq.~(\ref{strWC}) to the NDMIA, we first expand the functions around $\tilde m=0$ and apply the relations (\ref{rotrel}),
\eqn{
f(\tilde m_i^2,\tilde m_j^2,\dots)\cdot \Gamma_A^{iq_1*}\,\Gamma_B^{iq_1'}\cdot \Gamma_C^{jq_2*}\,\Gamma_D^{jq_2'}\cdots\ 
=\sum_{n_1,n_2,\dots} \frac{f^{(n_1,n_2,\dots)}(0)}{n_1!n_2!\cdots}\, [(M^2)^{n_1}]_{q_1q_1'}^{AB}\, [(M^2)^{n_2}]_{q_2q_2'}^{CD} \cdots\ ,
}
and then we split the mass matrix in two pieces: $M^2=X+\Delta$, where $X$ is diagonal and $\Delta\ll X$. Then we can resum the series in $X$, and keep only terms quadratic in $\Delta$. For example,
\eqa{
f(\tilde m_i^2,\tilde m_j^2,\tilde m_k^2)\cdot \Gamma_L^{is*}\,\Gamma_R^{ib}\cdot \Gamma_L^{js*}\,\Gamma_L^{jq}\cdot \Gamma_R^{kq*}\,\Gamma_L^{kb}=
F_{\tilde s_L}^{(2)}(X_{\tilde s_L},X_{\tilde b_R};X_{\tilde s_R},X_{\tilde b_L})\,\Delta_{sb}^{LR}\Delta_{sb}^{RL}\ ,\nn
}
with the function $F$ given by
\eqa{
&&\hspace{-0.5cm}F_{\tilde s_L}^{(2)}(X_{\tilde s_L},X_{\tilde b_R};X_{\tilde s_R},X_{\tilde b_L})=\nn\\[2mm]
&&\qquad\frac{
f(X_{\tilde s_L},X_{\tilde s_L},X_{\tilde s_R})-
f(X_{\tilde b_R},X_{\tilde s_L},X_{\tilde s_R})-
f(X_{\tilde s_L},X_{\tilde s_L},X_{\tilde b_L})+
f(X_{\tilde b_R},X_{\tilde s_L},X_{\tilde b_L})}{(X_{\tilde s_L}-X_{\tilde b_R})(X_{\tilde s_R}-X_{\tilde b_L})}\ .\nn
}
From the NDMIA results one can also recover the Wilson coefficients in the MIA by taking the limit in which all $X_{\tilde q}$'s are equal. In the non-degenerate case, the dimensionless mass insertions can be defined normalizing by any of the different squark mass parameters. It is customary to normalize by some ``average'' squark mass, although it makes no real difference as long as in the degenerate limit the definition coincides with the one in the MIA. We shall understand that they are normalized by the smaller diagonal entry of the squark mass matrix.

\section{Results}
\label{sec:res}

The full results for the LO and NLO Wilson coefficients in the $\rm \overline{MS}$-NDR scheme are presented in Appendix~\ref{sec:functions}. In this section we present some results for the NLO Wilson coefficients in a scenario with a hierarchy of masses, making a comparison with the MIA. In line with the rest of the paper, we keep focusing on the $B_s$ system for illustration.

The considered scenario is obtained from the mass insertion approximation with non-degenerate diagonal entries, taking a common mass for the first two generation squarks ($\tilde m_{12}$), different from a common mass for third generation squarks ($\tilde m_3$). This corresponds to the ``hierarchical'' scenario of Ref.~\cite{arXiv:0812.3610}, where $\tilde m_3$ is assumed to be near the electroweak scale, and $\tilde m_{12}$ is allowed to be heavy, up to several TeV. We therefore denote $x_h\equiv \tilde m_{12}^2/m_{\tilde g}^2$ and $x_l\equiv \tilde m_{3}^2/m_{\tilde g}^2$, for \emph{heavy} and \emph{light} respectively. Note that $\tilde m_{12}$ and $\tilde m_{3}$ are related to the true masses by corrections of $\op (\delta)$. In this case the mass insertions are normalized to $\tilde m_3$, which corresponds to the MIA definition when $x_h\to x_l$.

\begin{figure}
\begin{center}
\psfrag{xh}{$x_h$} \psfrag{C2NLO}{\hspace{-0.5cm}$\stackrel{1\ +\ C_2^{\rm \sss NLO}/C_2^{\rm \sss LO}}{}$}
\psfrag{C3NLO}{$\stackrel{\hspace{-0.5cm}1\ +\ C_3^{\rm \sss NLO}/C_3^{\rm \sss LO}}{}$}
\includegraphics[width=7cm]{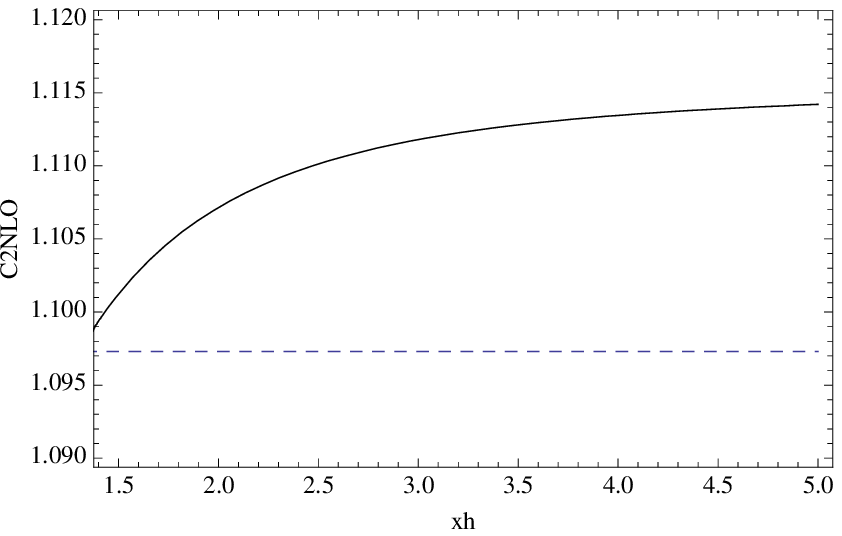}\qquad\quad
\includegraphics[width=7cm]{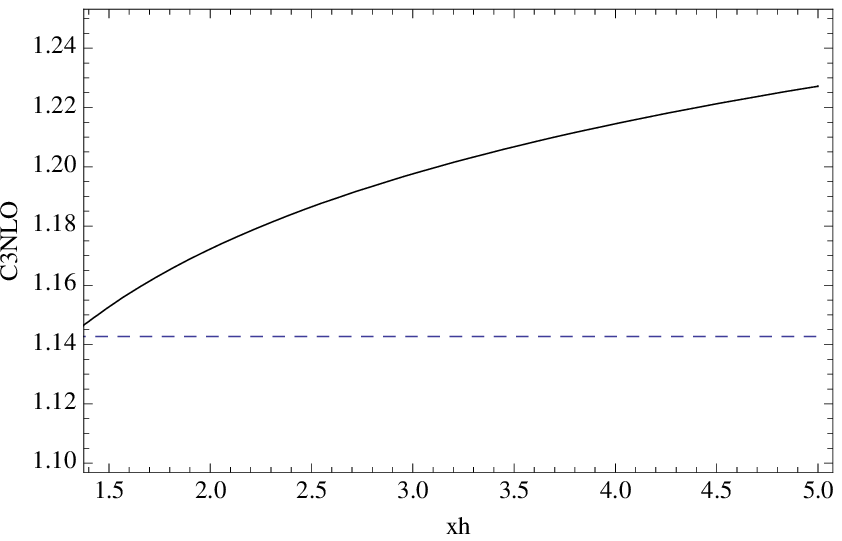}
\vspace{-0.5cm}
\end{center}
\caption{Relative importance of the full NLO result with respect to the LO, for $C_2$ and $C_3$, as a function of $x_h\equiv \tilde m_{12}^2/m_{\tilde g}^2$, where $\tilde m_{12}$ is a common mass for first and second generation squarks. The dashed lines correspond to the degenerate (MIA) scenario. We have chosen $m_{\tilde g}=\mu=350~{\rm GeV}$, and $400~{\rm GeV}$ for third generation squark masses.}
\label{C23NLO}
\end{figure}

The plots in Fig.~\ref{C23NLO} illustrate the relative importance of the NLO corrections, as a function of the mass splitting between the light and heavy squarks. The NLO correction is typically a $\sim 10\%$ effect in the degenerate case, but its importance increases with the mass splitting. For heavy squarks of about a TeV, the NLO contribution to $C_3$ can be up to a $\sim 25\%$ correction. However it should be mentioned that this is true in the  NDR scheme and could vary in other schemes. Also, this depends on the matching scale $\mu$, but a high sensitivity to $\mu$ would be related to large scale ambiguity at LO that is efficiently reduced at NLO. This means that the NLO contributions are either numerically important, or they achieve a considerable reduction of theoretical errors, and both situations are not easily disentangled.

\begin{figure}
\begin{center}
\psfrag{xl}{$x_l$} \psfrag{C1NLO}{\hspace{-0.5cm}\small $C_1^{\rm \sss NLO}\ {\scriptstyle (\times 10^{8})}$}
\psfrag{C4NLO}{\hspace{-0.5cm}\small $C_4^{\rm \sss NLO}\ {\scriptstyle (\times 10^{6})}$}
\psfrag{C5NLO}{\hspace{-0.5cm}\small $C_5^{\rm \sss NLO}\ {\scriptstyle (\times 10^{7})}$}
\includegraphics[width=4.9cm,height=4cm]{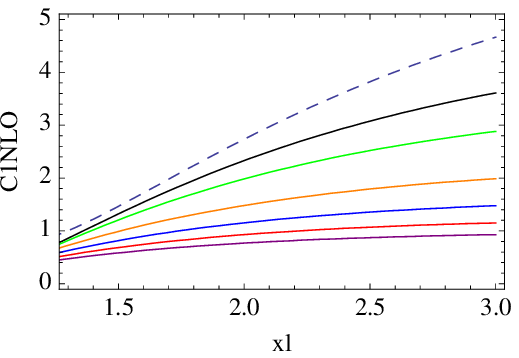}\quad
\includegraphics[width=4.9cm,height=4cm]{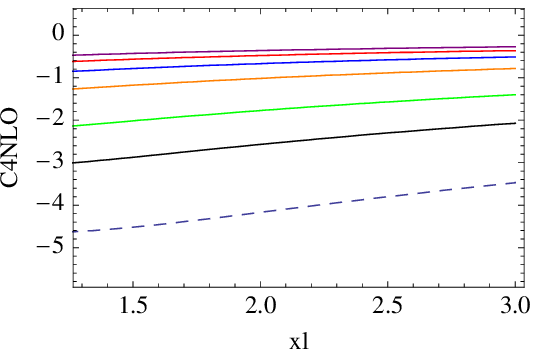}\quad
\includegraphics[width=4.9cm,height=4cm]{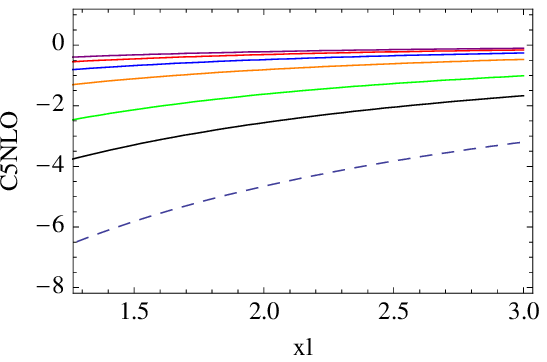}
\vspace{-0.5cm}
\end{center}
\caption{NLO Wilson coefficients $C_1^{NLO}$, $C_4^{NLO}$ and $C_5^{NLO}$ as a function of $x_l\equiv \tilde m_{3}^2/m_{\tilde g}^2$, where $\tilde m_{3}$ is a common mass for third generation squarks. The plots are in units of $(\alpha_s^3/\pi) \delta_{LL}^2$ and $(\alpha_s^3/\pi) \delta_{LL}\delta_{RR}$ for $C_1$ and $C_{4,5}$ respectively. The different lines correspond to different values of $x_h\equiv \tilde m_{12}^2/m_{\tilde g}^2$, and range from the MIA case, $x_h=x_l$ (dashed) to $x_h=1.5 x_l, 2 x_l,3 x_l, 4 x_l, 5 x_l, 6 x_l$.}
\label{C145NLO}
\end{figure}

In order to analyze more closely the role of the mass splittings in the NLO corrections, we consider $C_1, C_4$ and $C_5$ as a function $x_l$, for different splittings between $x_l$ and $x_h$. This is shown in Fig.~\ref{C145NLO}, where the dashed lines correspond to $x_l=x_h$ (that is, the degenerate case), and --departing smoothly from that limit-- the solid lines show increasing values of $x_h/x_l$. In these plots we take $\delta_{L,R}=\delta_{R,L}=0$ and $m_{\tilde g}=\mu=350~{\rm GeV}$. We see that increasing the heavy scale tends to reduce systematically the size of the NLO contribution, being largest in the degenerate case.

As mentioned before, this is scheme dependent (although the conclusion might be more general). In any case, the true impact of the NLO corrections can only be established by analyzing their effect on observables. A full phenomenological analysis of these corrections and their impact on the bounds on the mass insertions (beyond the mass insertion approximation) is worthwhile, and will be presented elsewhere.

\section{Conclusions}

In this paper we have presented the computation of the NLO strong interaction corrections to the Wilson coefficients relevant for $\Delta F=2$ processes in the MSSM, beyond the mass insertion approximation. The full results for the Wilson coefficients in the NDR scheme are given in Appendix~\ref{sec:functions}.

These results are relevant for two reasons. First, NLO corrections are necessary to cancel renormalization scheme and scale dependence from the renormalized operators. This has the effect of a considerable reduction in the theoretical error. Second, in order to study scenarios with significant mass splittings one must depart from the degenerate mass insertion approximation. We have shown some illustrative examples of the effect of mass splittings in the NLO Wilson coefficients as compared to the degenerate case.

A full phenomenological study of neutral meson mixing incorporating these new corrections will be presented in the future. Corresponding calculations for $\Delta F=1$ processes are underway, and when available will allow to perform a complete analysis of correlations between decay and mixing observables of neutral mesons at NLO in $\alpha_s$.

\subsubsection*{Acknowledgements}

I would like to thank Luca Silvestrini, Enrico Franco and Miguel Nebot for useful discussions, as well as Ulrich Nierste, Andreas Crivellin and  Diego Guadagnoli for correspondence. J.V. is associated to the Dipartimento di Fisica, Universit\`a di Roma `La Sapienza'.

\appendix

\section{List of diagrams}
\label{app1}

In this appendix we present the set of one- and two-loop feynman diagrams that contribute to the MSSM amplitude. We focus on the case of $B_s$ mixing; for other neutral meson systems the external quarks must be changed. Dashed lines and $i,j,k,l$ denote squarks, solid arrowed lines and $q$ denote quarks and solid lines without arrows are (majorana) gluinos. We show all the possible topologies with the different insertions of external quark flavors, but an additional multiplicity is present interchanging initial and final external quarks (whenever they correspond to \emph{different} contractions). For example, at leading order there are two diagrams of each of the two topologies shown in Appendix~\ref{appLO}, with appropriate signs given by a reference order of external legs.

\subsection{LO diagrams}
\label{appLO}

\begin{center}
\psfrag{s}{$s$} \psfrag{b}{$b$} \psfrag{i}{$i$} \psfrag{j}{$j$}
\includegraphics[height=3.5cm]{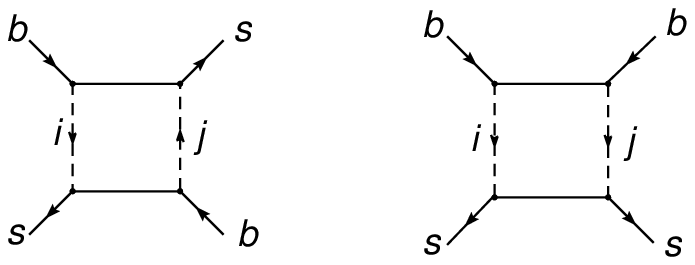}
\end{center}

\subsection{IR divergent diagrams}
\label{appIR}

\begin{center}
\psfrag{s}{$s$} \psfrag{b}{$b$} \psfrag{i}{$i$} \psfrag{j}{$j$}
\includegraphics[width=16cm,height=6cm]{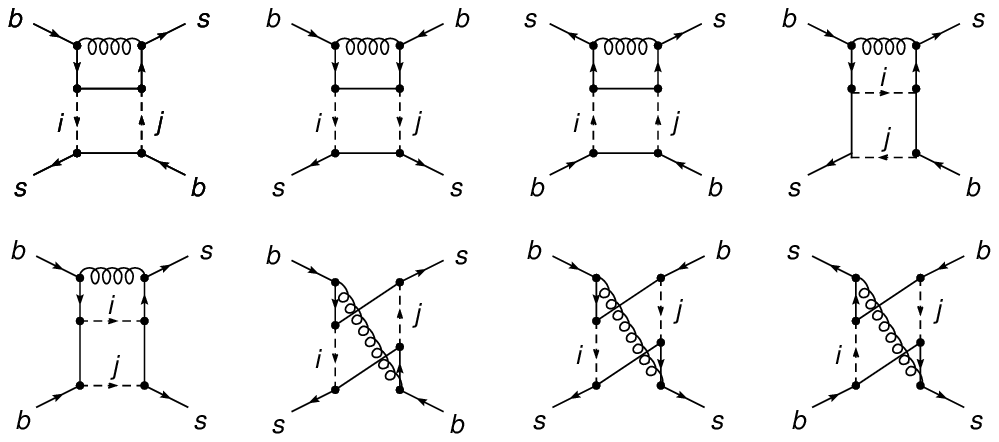}
\end{center}

\subsection{UV divergent diagrams}
\label{appUV}

\subsubsection{Vertex corrections}
\label{appVC}

\begin{center}
\psfrag{s}{$s$} \psfrag{b}{$b$} \psfrag{i}{$i$} \psfrag{j}{$j$}
\includegraphics[width=16cm,height=9cm]{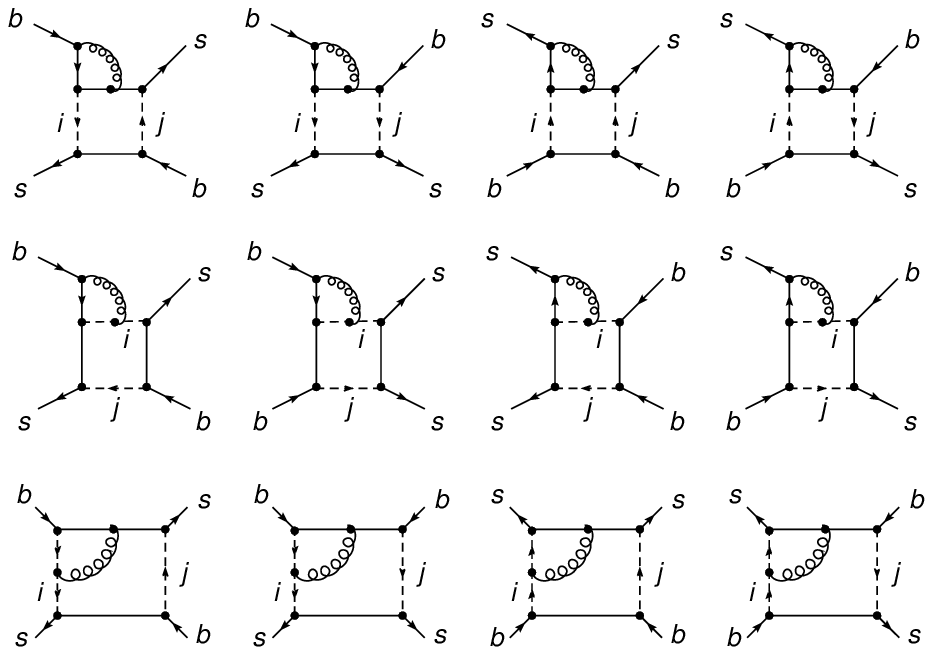}
\end{center}

\subsubsection{Gluino self-energies}
\label{appGSE}

\begin{center}
\psfrag{s}{$s$} \psfrag{b}{$b$} \psfrag{i}{$i$} \psfrag{j}{$j$}
\psfrag{k}{$k$} \psfrag{l}{$l$} \psfrag{q}{$q$} 
\includegraphics[width=12cm,height=8cm]{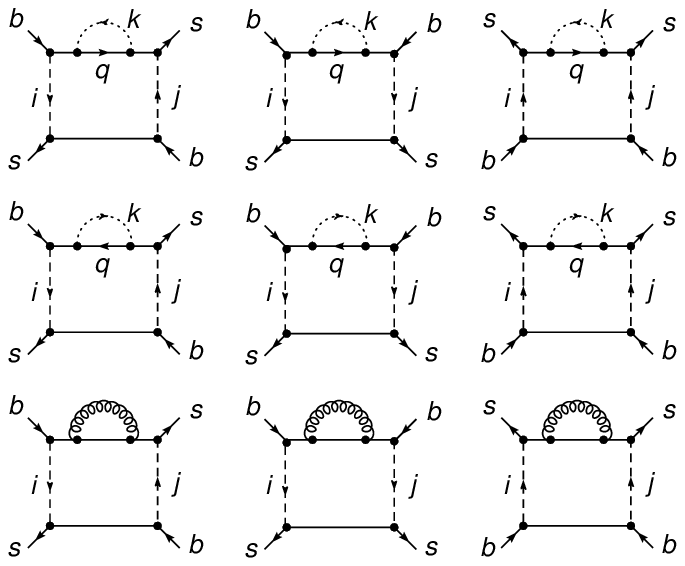}
\end{center}

\subsubsection{Squark self-energies}
\label{appSSE}

\begin{center}
\psfrag{s}{$s$} \psfrag{b}{$b$} \psfrag{i}{$i$} \psfrag{j}{$j$}
\psfrag{k}{$k$} \psfrag{l}{$l$} \psfrag{q}{$q$} 
\includegraphics[width=16cm,height=5cm]{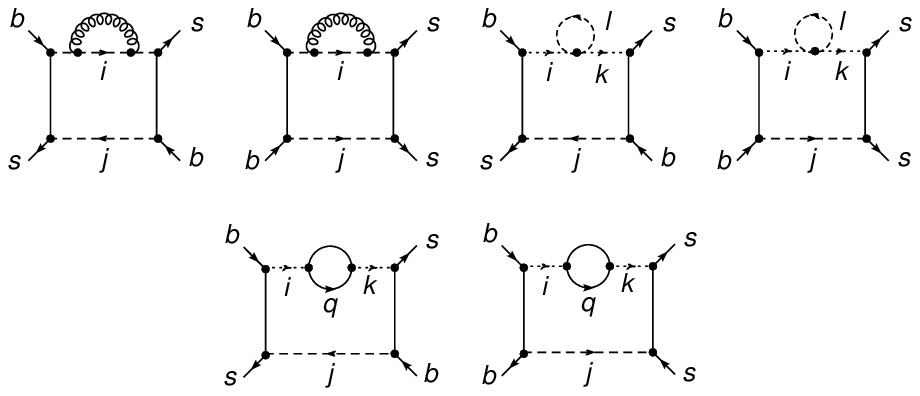}
\end{center}



\subsection{Finite NLO diagrams}
\label{appFIN}

\begin{center}
\psfrag{s}{$s$} \psfrag{b}{$b$} \psfrag{i}{$i$} \psfrag{j}{$j$}
\psfrag{k}{$k$} \psfrag{l}{$l$} \psfrag{q}{$q$} 
\includegraphics[width=16cm,height=8cm]{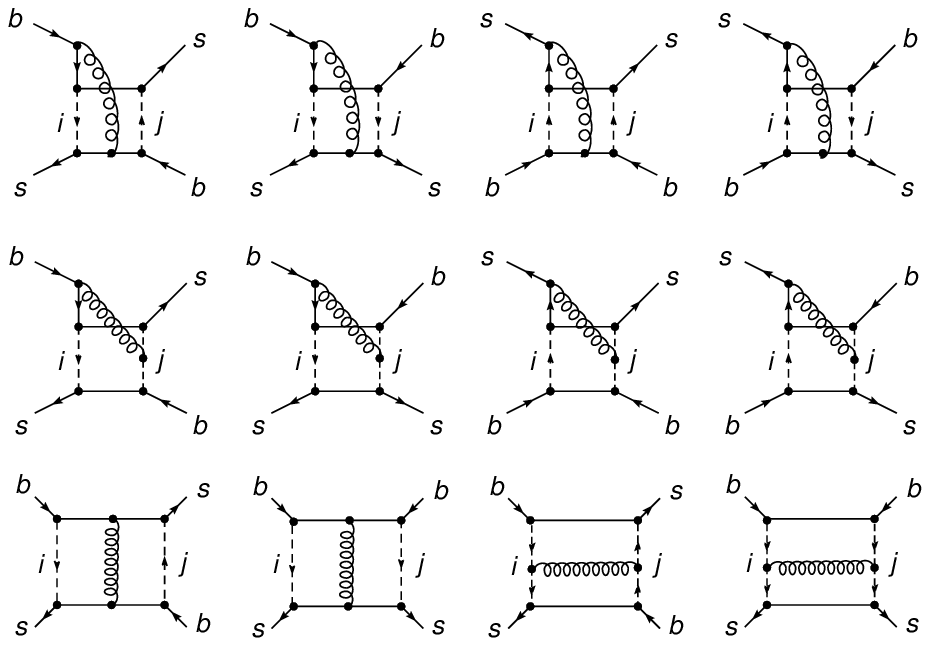}\\[0.5cm]
\includegraphics[width=16cm,height=5cm]{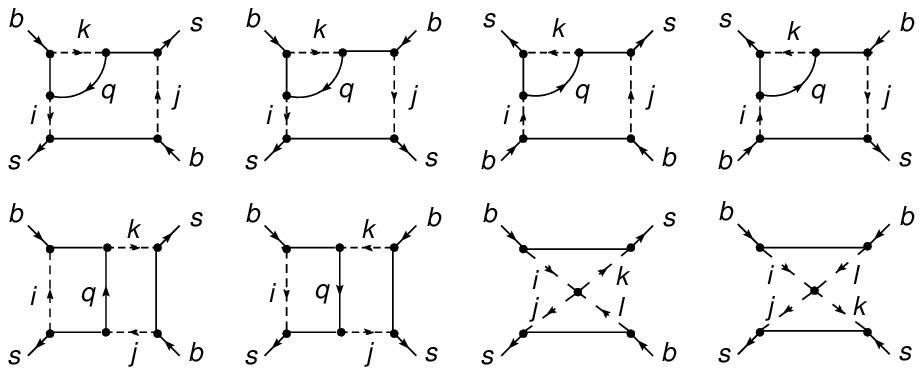}
\end{center}


\section{Wilson coefficients}
\label{sec:functions}

In this appendix we present the complete expressions for the Wilson coefficients at NLO in the $\rm \overline MS$-NDR scheme. We omit the results for $\tilde C_{1,2,3}$, which are obtained from $C_{1,2,3}$ by exchanging $L \leftrightarrow R$ in the squark rotation matrices. The coefficients are split into LO and NLO:
\eq{C_i(\mu)=C_i^{(0)}(\mu)+C_i^{(1)}(\mu)\ ,}
and depend on the matching scale $\mu$ through $\alpha_s$, $m_{\tilde g}$ and $\tilde m_i$, and explicitly through $\log(m_{\tilde g}^2/\mu^2)$.  The functions depend on the squark masses through the variables $x_i$ defined as $x_i\equiv \tilde m_i^2/m_{\tilde g}^2$.

The LO Wilson coefficients are given by

{
\eqa{
C_1^{(0)}(\mu)&=&\frac{\alpha_s^2}{12^2 m_{\tilde g}^2}\ \Big( 8\,G^{ij}-22\,H^{ij} \Big)\,\Gamma_L^{is*}\Gamma_L^{ib}\,\Gamma_L^{js*}\Gamma_L^{jb}\nn\\[2mm]
C_2^{(0)}(\mu)&=&\frac{\alpha_s^2}{12^2 m_{\tilde g}^2}\ 68\,G^{ij} \,\Gamma_R^{is*}\Gamma_L^{ib}\,\Gamma_R^{js*}\Gamma_L^{jb}\nn\\[2mm]
C_3^{(0)}(\mu)&=&-\frac{\alpha_s^2}{12^2 m_{\tilde g}^2}\ 12\,G^{ij} \,\Gamma_R^{is*}\Gamma_L^{ib}\,\Gamma_R^{js*}\Gamma_L^{jb}\nn\\[2mm]
C_4^{(0)}(\mu)&=&\frac{\alpha_s^2}{12^2 m_{\tilde g}^2}\bigg[  
\Big( 168\,G^{ij}+24\,H^{ij} \Big)\Gamma_R^{is*}\Gamma_R^{ib}\,\Gamma_L^{js*}\Gamma_L^{jb}
+44\,H^{ij}\,\Gamma_L^{is*}\Gamma_R^{ib}\,\Gamma_R^{js*}\Gamma_L^{jb}   \bigg]\nn\\[2mm]
C_5^{(0)}(\mu)&=&\frac{\alpha_s^2}{12^2 m_{\tilde g}^2}\bigg[  
\Big( 8\,G^{ij}-40\,H^{ij} \Big)\Gamma_R^{is*}\Gamma_R^{ib}\,\Gamma_L^{js*}\Gamma_L^{jb}
+ 60\,H^{ij} \,\Gamma_L^{is*}\Gamma_R^{ib}\,\Gamma_R^{js*}\Gamma_L^{jb}   \bigg]
\label{LOWC}}
}

where a sum is understood over $i,j=\tilde d_L, \tilde s_L, \tilde b_L, \tilde d_R, \tilde s_R, \tilde b_R$. The functions $G, H$ are:

{
\eqn{
G^{ij}=\frac{2x_i\log{x_i}-x_i^2+1}{(x_i-1)^2(x_i-x_j)}+(x_i\leftrightarrow x_j)\ ;\quad
H^{ij}=\frac{-2x_i^2\log{x_i}+3x_i^2-4x_i+1}{(x_i-1)^2(x_i-x_j)}+(x_i\leftrightarrow x_j) 
}
}

The NLO coefficient functions are the main result of this paper. They are scheme dependent; we present the results in the $\rm \overline MS$-NDR scheme, but can be translated easily to other schemes using formulae analogous to that presented in Ref~\cite{hep-ph/0606197}. The function $\Li(x)$ denotes the dilogarithm defined in the usual way:
\eq{\Li(x)=-\int_0^x dt\ \frac{\log(1-t)}{t}\ .}
Appropriate sums over indices are understood, in particular $q,q'=d,s,b$ and $Q=u,d,s,c,b,t$, and $i,j,k$ run over left and right-handed squarks of $u$ or $d$-type depending on the quark they appear with in the rotation matrix: for example a term containing $\Gamma_L^{ku}$ contains a sum over $k=\tilde u_L, \tilde c_L, \tilde t_L, \tilde u_R, \tilde c_R, \tilde t_R$. All dependence on $u$-type quarks and squarks come only from diagrams with gluino self-energies.  The NLO Wilson coefficients read

{
\eqa{
&&C_1^{(1)}(\mu) =\frac{\alpha_s^3}{12^3\pi m_{\tilde g}^2} \Big[ (f_1^{ij}+f_1^{ji})\,\Gamma_L^{is*}\Gamma_L^{ib}\,\Gamma_L^{js*}\Gamma_L^{jb}
+ g_{1,1}^{ijk}\,\Gamma_L^{is*}\Gamma_L^{ib}\, \Gamma_L^{js*}\Gamma_L^{jq}\,\Gamma_L^{kq*}\Gamma_L^{kb}\nn\\
&&\qquad + g_{1,2}^{ijk}\,\Gamma_L^{is*}\Gamma_L^{ib}\, \Gamma_L^{js*}\Gamma_R^{jq}\,\Gamma_R^{kq*}\Gamma_L^{kb}
+ h_1^{ijk}\,\Gamma_L^{is*}\Gamma_L^{ib}\, \Gamma_L^{js*}\Gamma_L^{jb}\,(\Gamma_L^{kQ*}\Gamma_L^{kQ}+\Gamma_R^{kQ*}\Gamma_R^{kQ})\nn\\
&&\qquad+ m_1^{ijk}\,\Gamma_L^{is*}\Gamma_L^{ib}\, \Gamma_L^{js*}\Gamma_L^{jb}\,(\Gamma_L^{ks*}\Gamma_L^{ks}+\Gamma_L^{kb*}\Gamma_L^{kb})\nn\\
&&\qquad+ n_1^{ijkl}\,(\Gamma_L^{iq*}\Gamma_L^{ib}\, \Gamma_L^{js*}\Gamma_L^{jb}\,\Gamma_L^{ks*}\Gamma_L^{kq'}\,\Gamma_L^{lq'*}\Gamma_L^{lq}+
\Gamma_R^{iq*}\Gamma_L^{ib}\, \Gamma_L^{js*}\Gamma_L^{jb}\,\Gamma_L^{ks*}\Gamma_R^{kq'}\,\Gamma_R^{lq'*}\Gamma_R^{lq})\nn\\
&&\qquad+p_1^{ijkl}\,(\Gamma_L^{iq*}\Gamma_L^{ib}\, \Gamma_L^{js*}\Gamma_L^{jq}\,\Gamma_L^{ks*}\Gamma_L^{kq'}\,\Gamma_L^{lq'*}\Gamma_L^{lb}
-\Gamma_L^{iq*}\Gamma_L^{ib}\, \Gamma_L^{js*}\Gamma_L^{jq}\,\Gamma_L^{ks*}\Gamma_R^{kq'}\,\Gamma_R^{lq'*}\Gamma_L^{lb}\nn\\
&&\qquad-\Gamma_R^{iq*}\Gamma_L^{ib}\, \Gamma_L^{js*}\Gamma_R^{jq}\,\Gamma_L^{ks*}\Gamma_L^{kq'}\,\Gamma_L^{lq'*}\Gamma_L^{lb}
+\Gamma_R^{iq*}\Gamma_L^{ib}\, \Gamma_L^{js*}\Gamma_R^{jq}\,\Gamma_L^{ks*}\Gamma_R^{kq'}\,\Gamma_R^{lq'*}\Gamma_L^{lb})
\Big]\\[2mm]
&&C_2^{(1)}(\mu) = \frac{\alpha_s^3}{12^3\pi m_{\tilde g}^2} \Big[ (f_2^{ij}+f_2^{ji})\,\Gamma_R^{is*}\Gamma_L^{ib}\,\Gamma_R^{js*}\Gamma_L^{jb}
+ g_2^{ijk}\,\Gamma_R^{is*}\Gamma_L^{ib}\, \Gamma_R^{js*}\Gamma_L^{jq}\,\Gamma_L^{kq*}\Gamma_L^{kb}\nn\\
&&\qquad+ g_2^{ikj}\,\Gamma_R^{is*}\Gamma_L^{ib}\, \Gamma_R^{js*}\Gamma_R^{jq}\,\Gamma_R^{kq*}\Gamma_L^{kb}
+ h_2^{ijk}\,\Gamma_R^{is*}\Gamma_L^{ib}\, \Gamma_R^{js*}\Gamma_L^{jb}\,(\Gamma_L^{kQ*}\Gamma_L^{kQ}+\Gamma_R^{kQ*}\Gamma_R^{kQ})\nn\\
&&\qquad+ m_2^{ijk}\,\Gamma_R^{is*}\Gamma_L^{ib}\, \Gamma_R^{js*}\Gamma_L^{jb}\,(\Gamma_R^{ks*}\Gamma_R^{ks}+\Gamma_L^{kb*}\Gamma_L^{kb})\nn\\
&&\qquad+ n_2^{ijkl}\,(\Gamma_L^{iq*}\Gamma_L^{ib}\, \Gamma_R^{js*}\Gamma_L^{jb}\,\Gamma_R^{ks*}\Gamma_L^{kq'}\,\Gamma_L^{lq'*}\Gamma_L^{lq}+
\Gamma_R^{iq*}\Gamma_L^{ib}\, \Gamma_R^{js*}\Gamma_L^{jb}\,\Gamma_R^{ks*}\Gamma_R^{kq'}\,\Gamma_R^{lq'*}\Gamma_R^{lq})\nn\\
&&\qquad+p_2^{ijkl}\,(\Gamma_L^{iq*}\Gamma_L^{ib}\, \Gamma_R^{js*}\Gamma_L^{jq}\,\Gamma_R^{ks*}\Gamma_L^{kq'}\,\Gamma_L^{lq'*}\Gamma_L^{lb}
+\Gamma_R^{iq*}\Gamma_L^{ib}\, \Gamma_R^{js*}\Gamma_R^{jq}\,\Gamma_R^{ks*}\Gamma_R^{kq'}\,\Gamma_R^{lq'*}\Gamma_L^{lb}\nn\\
&&\qquad-\Gamma_L^{iq*}\Gamma_L^{ib}\, \Gamma_R^{js*}\Gamma_L^{jq}\,\Gamma_R^{ks*}\Gamma_R^{kq'}\,\Gamma_R^{lq'*}\Gamma_L^{lb}
-\Gamma_R^{iq*}\Gamma_L^{ib}\, \Gamma_R^{js*}\Gamma_R^{jq}\,\Gamma_R^{ks*}\Gamma_L^{kq'}\,\Gamma_L^{lq'*}\Gamma_L^{lb})
\Big]\\[2mm]
&&C_3^{(1)}(\mu) =\frac{\alpha_s^3}{12^3\pi m_{\tilde g}^2} \Big[  (f_3^{ij}+f_3^{ji})\,\Gamma_R^{is*}\Gamma_L^{ib}\,\Gamma_R^{js*}\Gamma_L^{jb}
+ g_3^{ijk}\,\Gamma_R^{is*}\Gamma_L^{ib}\, \Gamma_R^{js*}\Gamma_L^{jq}\,\Gamma_L^{kq*}\Gamma_L^{kb}\nn\\
&&\qquad+ g_3^{ikj}\,\Gamma_R^{is*}\Gamma_L^{ib}\, \Gamma_R^{js*}\Gamma_R^{jq}\,\Gamma_R^{kq*}\Gamma_L^{kb}
+ h_3^{ijk}\,\Gamma_R^{is*}\Gamma_L^{ib}\, \Gamma_R^{js*}\Gamma_L^{jb}\,(\Gamma_L^{kQ*}\Gamma_L^{kQ}+\Gamma_R^{kQ*}\Gamma_R^{kQ})\nn\\
&&\qquad+ m_3^{ijk}\,\Gamma_R^{is*}\Gamma_L^{ib}\, \Gamma_R^{js*}\Gamma_L^{jb}\,(\Gamma_R^{ks*}\Gamma_R^{ks}+\Gamma_L^{kb*}\Gamma_L^{kb})\nn\\
&&\qquad+ n_3^{ijkl}\,(\Gamma_L^{iq*}\Gamma_L^{ib}\, \Gamma_R^{js*}\Gamma_L^{jb}\,\Gamma_R^{ks*}\Gamma_L^{kq'}\,\Gamma_L^{lq'*}\Gamma_L^{lq}+
\Gamma_R^{iq*}\Gamma_L^{ib}\, \Gamma_R^{js*}\Gamma_L^{jb}\,\Gamma_R^{ks*}\Gamma_R^{kq'}\,\Gamma_R^{lq'*}\Gamma_R^{lq})\nn\\
&&\qquad+p_3^{ijkl}\,(\Gamma_L^{iq*}\Gamma_L^{ib}\, \Gamma_R^{js*}\Gamma_L^{jq}\,\Gamma_R^{ks*}\Gamma_L^{kq'}\,\Gamma_L^{lq'*}\Gamma_L^{lb}
+\Gamma_R^{iq*}\Gamma_L^{ib}\, \Gamma_R^{js*}\Gamma_R^{jq}\,\Gamma_R^{ks*}\Gamma_R^{kq'}\,\Gamma_R^{lq'*}\Gamma_L^{lb}\nn\\
&&\qquad-\Gamma_L^{iq*}\Gamma_L^{ib}\, \Gamma_R^{js*}\Gamma_L^{jq}\,\Gamma_R^{ks*}\Gamma_R^{kq'}\,\Gamma_R^{lq'*}\Gamma_L^{lb}
-\Gamma_R^{iq*}\Gamma_L^{ib}\, \Gamma_R^{js*}\Gamma_R^{jq}\,\Gamma_R^{ks*}\Gamma_L^{kq'}\,\Gamma_L^{lq'*}\Gamma_L^{lb})
\Big]\\[2mm]
&&C_4^{(1)}(\mu) = \frac{\alpha_s^3}{12^3\pi m_{\tilde g}^2} \Big[ (f_{4,1}^{ij}+f_{4,1}^{ji})\,\Gamma_L^{is*}\Gamma_L^{ib}\,\Gamma_R^{js*}\Gamma_R^{jb}+
 (f_{4,2}^{ij}+f_{4,2}^{ji})\,\Gamma_L^{is*}\Gamma_R^{ib}\,\Gamma_R^{js*}\Gamma_L^{jb}\nn\\
&&\qquad+ g_{4,1}^{ijk}\,(\Gamma_L^{is*}\Gamma_R^{ib}\, \Gamma_R^{js*}\Gamma_L^{jq}\,\Gamma_L^{kq*}\Gamma_L^{kb}+(L\leftrightarrow R))
+ g_{4,2}^{ijk}\,(\Gamma_L^{is*}\Gamma_L^{ib}\, \Gamma_R^{js*}\Gamma_L^{jq}\,\Gamma_L^{kq*}\Gamma_R^{kb}+(L\leftrightarrow R))\nn\\
&&\qquad+ g_{4,3}^{ijk}\,(\Gamma_L^{is*}\Gamma_L^{ib}\, \Gamma_R^{js*}\Gamma_R^{jq}\,\Gamma_R^{kq*}\Gamma_R^{kb}+(L\leftrightarrow R))
+ g_{4,4}^{ijk}\,(\Gamma_L^{is*}\Gamma_R^{ib}\, \Gamma_R^{js*}\Gamma_R^{jq}\,\Gamma_R^{kq*}\Gamma_L^{kb}+(L\leftrightarrow R))\nn\\
&&\qquad+ h_{4,1}^{ijk}\,\Gamma_L^{is*}\Gamma_L^{ib}\, \Gamma_R^{js*}\Gamma_R^{jb}\,(\Gamma_L^{kQ*}\Gamma_L^{kQ}+\Gamma_R^{kQ*}\Gamma_R^{kQ})
+ h_{4,2}^{ijk}\,\Gamma_R^{is*}\Gamma_R^{ib}\, \Gamma_L^{js*}\Gamma_L^{jb}\,(\Gamma_L^{kQ*}\Gamma_L^{kQ}+\Gamma_R^{kQ*}\Gamma_R^{kQ})\nn\\
&&\qquad+ h_{4,3}^{ijk}\,(\Gamma_L^{is*}\Gamma_R^{ib}\, \Gamma_R^{js*}\Gamma_L^{jb}\,(\Gamma_L^{kQ*}\Gamma_L^{kQ}+\Gamma_R^{kQ*}\Gamma_R^{kQ}) +(L\leftrightarrow R))\nn\\
&&\qquad+ m_{4,1}^{ijk}\,(\Gamma_L^{is*}\Gamma_L^{ib}\, \Gamma_R^{js*}\Gamma_R^{jb}\,(\Gamma_L^{ks*}\Gamma_L^{ks}+\Gamma_L^{kb*}\Gamma_L^{kb}) +(L\leftrightarrow R))\nn\\
&&\qquad+ m_{4,2}^{ijk}\,(\Gamma_L^{is*}\Gamma_R^{ib}\, \Gamma_R^{js*}\Gamma_L^{jb}\,(\Gamma_L^{ks*}\Gamma_L^{ks}+\Gamma_R^{kb*}\Gamma_R^{kb}) +(L\leftrightarrow R))\nn\\
&&\qquad+ n_{4,1}^{ijkl}\,(\Gamma_L^{iq*}\Gamma_L^{ib}\, \Gamma_R^{js*}\Gamma_R^{jb}\,\Gamma_L^{ks*}\Gamma_L^{kq'}\,\Gamma_L^{lq'*}\Gamma_L^{lq}+
\Gamma_L^{iq*}\Gamma_R^{ib}\, \Gamma_L^{js*}\Gamma_L^{jb}\,\Gamma_R^{ks*}\Gamma_L^{kq'}\,\Gamma_L^{lq'*}\Gamma_L^{lq}+(L\leftrightarrow R))\nn\\
&&\qquad+ n_{4,2}^{ijkl}\,(\Gamma_L^{iq*}\Gamma_L^{ib}\, \Gamma_L^{js*}\Gamma_R^{jb}\,\Gamma_R^{ks*}\Gamma_L^{kq'}\,\Gamma_L^{lq'*}\Gamma_L^{lq}+
\Gamma_L^{iq*}\Gamma_R^{ib}\, \Gamma_R^{js*}\Gamma_L^{jb}\,\Gamma_L^{ks*}\Gamma_L^{kq'}\,\Gamma_L^{lq'*}\Gamma_L^{lq}+(L\leftrightarrow R))\nn\\
&&\qquad+p_{4,1}^{ijkl}\,(\Gamma_L^{iq*}\Gamma_L^{ib}\, \Gamma_R^{js*}\Gamma_L^{jq}\,\Gamma_L^{ks*}\Gamma_L^{kq'}\,\Gamma_L^{lq'*}\Gamma_R^{lb}
-\Gamma_L^{iq*}\Gamma_L^{ib}\, \Gamma_R^{js*}\Gamma_L^{jq}\,\Gamma_L^{ks*}\Gamma_R^{kq'}\,\Gamma_R^{lq'*}\Gamma_R^{lb}\nn\\
&&\qquad-\Gamma_R^{iq*}\Gamma_L^{ib}\, \Gamma_R^{js*}\Gamma_R^{jq}\,\Gamma_L^{ks*}\Gamma_L^{kq'}\,\Gamma_L^{lq'*}\Gamma_R^{lb}
+\Gamma_R^{iq*}\Gamma_L^{ib}\, \Gamma_R^{js*}\Gamma_R^{jq}\,\Gamma_L^{ks*}\Gamma_R^{kq'}\,\Gamma_R^{lq'*}\Gamma_R^{lb})\nn\\
&&\qquad+p_{4,2}^{ijkl}\,(\Gamma_L^{iq*}\Gamma_L^{ib}\, \Gamma_L^{js*}\Gamma_L^{jq}\,\Gamma_R^{ks*}\Gamma_L^{kq'}\,\Gamma_L^{lq'*}\Gamma_R^{lb}
-\Gamma_L^{iq*}\Gamma_L^{ib}\, \Gamma_L^{js*}\Gamma_L^{jq}\,\Gamma_R^{ks*}\Gamma_R^{kq'}\,\Gamma_R^{lq'*}\Gamma_R^{lb}\nn\\
&&\qquad-\Gamma_R^{iq*}\Gamma_L^{ib}\, \Gamma_L^{js*}\Gamma_R^{jq}\,\Gamma_R^{ks*}\Gamma_L^{kq'}\,\Gamma_L^{lq'*}\Gamma_R^{lb}
+\Gamma_R^{iq*}\Gamma_L^{ib}\, \Gamma_L^{js*}\Gamma_R^{jq}\,\Gamma_R^{ks*}\Gamma_R^{kq'}\,\Gamma_R^{lq'*}\Gamma_R^{lb})
\Big]\\[2mm]
&&C_5^{(1)}(\mu) = \frac{\alpha_s^3}{12^3\pi m_{\tilde g}^2} \Big[ (f_{5,1}^{ij}+f_{5,1}^{ji})\,\Gamma_L^{is*}\Gamma_L^{ib}\,\Gamma_R^{js*}\Gamma_R^{jb}+
 (f_{5,2}^{ij}+f_{5,2}^{ji})\,\Gamma_L^{is*}\Gamma_R^{ib}\,\Gamma_R^{js*}\Gamma_L^{jb}\nn\\
&&\qquad+ g_{5,1}^{ijk}\,(\Gamma_L^{is*}\Gamma_R^{ib}\, \Gamma_R^{js*}\Gamma_L^{jq}\,\Gamma_L^{kq*}\Gamma_L^{kb}+(L\leftrightarrow R))
+ g_{5,2}^{ijk}\,(\Gamma_L^{is*}\Gamma_L^{ib}\, \Gamma_R^{js*}\Gamma_L^{jq}\,\Gamma_L^{kq*}\Gamma_R^{kb}+(L\leftrightarrow R))\nn\\
&&\qquad+ g_{5,3}^{ijk}\,(\Gamma_L^{is*}\Gamma_L^{ib}\, \Gamma_R^{js*}\Gamma_R^{jq}\,\Gamma_R^{kq*}\Gamma_R^{kb}+(L\leftrightarrow R))
+ g_{5,4}^{ijk}\,(\Gamma_L^{is*}\Gamma_R^{ib}\, \Gamma_R^{js*}\Gamma_R^{jq}\,\Gamma_R^{kq*}\Gamma_L^{kb}+(L\leftrightarrow R))\nn\\
&&\qquad+ h_{5,1}^{ijk}\,\Gamma_L^{is*}\Gamma_L^{ib}\, \Gamma_R^{js*}\Gamma_R^{jb}\,(\Gamma_L^{kQ*}\Gamma_L^{kQ}+\Gamma_R^{kQ*}\Gamma_R^{kQ})
+ h_{5,2}^{ijk}\,\Gamma_R^{is*}\Gamma_R^{ib}\, \Gamma_L^{js*}\Gamma_L^{jb}\,(\Gamma_L^{kQ*}\Gamma_L^{kQ}+\Gamma_R^{kQ*}\Gamma_R^{kQ})\nn\\
&&\qquad+ h_{5,3}^{ijk}\,(\Gamma_L^{is*}\Gamma_R^{ib}\, \Gamma_R^{js*}\Gamma_L^{jb}\,(\Gamma_L^{kQ*}\Gamma_L^{kQ}+\Gamma_R^{kQ*}\Gamma_R^{kQ}) +(L\leftrightarrow R))\nn\\
&&\qquad+ m_{5,1}^{ijk}\,(\Gamma_L^{is*}\Gamma_L^{ib}\, \Gamma_R^{js*}\Gamma_R^{jb}\,(\Gamma_L^{ks*}\Gamma_L^{ks}+\Gamma_L^{kb*}\Gamma_L^{kb}) +(L\leftrightarrow R))\nn\\
&&\qquad+ m_{5,2}^{ijk}\,(\Gamma_L^{is*}\Gamma_R^{ib}\, \Gamma_R^{js*}\Gamma_L^{jb}\,(\Gamma_L^{ks*}\Gamma_L^{ks}+\Gamma_R^{kb*}\Gamma_R^{kb}) +(L\leftrightarrow R))\nn\\
&&\qquad+ n_{5,1}^{ijkl}\,(\Gamma_L^{iq*}\Gamma_L^{ib}\, \Gamma_R^{js*}\Gamma_R^{jb}\,\Gamma_L^{ks*}\Gamma_L^{kq'}\,\Gamma_L^{lq'*}\Gamma_L^{lq}+
\Gamma_L^{iq*}\Gamma_R^{ib}\, \Gamma_L^{js*}\Gamma_L^{jb}\,\Gamma_R^{ks*}\Gamma_L^{kq'}\,\Gamma_L^{lq'*}\Gamma_L^{lq}+(L\leftrightarrow R))\nn\\
&&\qquad+ n_{5,2}^{ijkl}\,(\Gamma_L^{iq*}\Gamma_L^{ib}\, \Gamma_L^{js*}\Gamma_R^{jb}\,\Gamma_R^{ks*}\Gamma_L^{kq'}\,\Gamma_L^{lq'*}\Gamma_L^{lq}+
\Gamma_L^{iq*}\Gamma_R^{ib}\, \Gamma_R^{js*}\Gamma_L^{jb}\,\Gamma_L^{ks*}\Gamma_L^{kq'}\,\Gamma_L^{lq'*}\Gamma_L^{lq}+(L\leftrightarrow R))\nn\\
&&\qquad+p_{5,1}^{ijkl}\,(\Gamma_L^{iq*}\Gamma_L^{ib}\, \Gamma_R^{js*}\Gamma_L^{jq}\,\Gamma_L^{ks*}\Gamma_L^{kq'}\,\Gamma_L^{lq'*}\Gamma_R^{lb}
-\Gamma_L^{iq*}\Gamma_L^{ib}\, \Gamma_R^{js*}\Gamma_L^{jq}\,\Gamma_L^{ks*}\Gamma_R^{kq'}\,\Gamma_R^{lq'*}\Gamma_R^{lb}\nn\\
&&\qquad-\Gamma_R^{iq*}\Gamma_L^{ib}\, \Gamma_R^{js*}\Gamma_R^{jq}\,\Gamma_L^{ks*}\Gamma_L^{kq'}\,\Gamma_L^{lq'*}\Gamma_R^{lb}
+\Gamma_R^{iq*}\Gamma_L^{ib}\, \Gamma_R^{js*}\Gamma_R^{jq}\,\Gamma_L^{ks*}\Gamma_R^{kq'}\,\Gamma_R^{lq'*}\Gamma_R^{lb})\nn\\
&&\qquad+p_{5,2}^{ijkl}\,(\Gamma_L^{iq*}\Gamma_L^{ib}\, \Gamma_L^{js*}\Gamma_L^{jq}\,\Gamma_R^{ks*}\Gamma_L^{kq'}\,\Gamma_L^{lq'*}\Gamma_R^{lb}
-\Gamma_L^{iq*}\Gamma_L^{ib}\, \Gamma_L^{js*}\Gamma_L^{jq}\,\Gamma_R^{ks*}\Gamma_R^{kq'}\,\Gamma_R^{lq'*}\Gamma_R^{lb}\nn\\
&&\qquad-\Gamma_R^{iq*}\Gamma_L^{ib}\, \Gamma_L^{js*}\Gamma_R^{jq}\,\Gamma_R^{ks*}\Gamma_L^{kq'}\,\Gamma_L^{lq'*}\Gamma_R^{lb}
+\Gamma_R^{iq*}\Gamma_L^{ib}\, \Gamma_L^{js*}\Gamma_R^{jq}\,\Gamma_R^{ks*}\Gamma_R^{kq'}\,\Gamma_R^{lq'*}\Gamma_R^{lb})
\Big]
}
}

The coefficients $g^{ijk}, h^{ijk}, m^{ijk}, n^{ijkl}$ and $p^{ijkl}$ are the following,

{
\eqa{
g_{1,1}^{ijk}&=&15(A_1^{ijk}+A_1^{jik})-25 A_5^{ijk}+34 A_6^{ijk} + (j\leftrightarrow k)\nn\\
g_{1,2}^{ijk}&=&-8(A_3^{ijk}+A_3^{jik})+81(A_2^{ijk}+A_2^{jik})+34 A_5^{ijk}-25 A_6^{ijk} + (j\leftrightarrow k)\nn\\
g_2^{ijk}&=&34(A_1^{ijk}+A_1^{jik}-2A_3^{ikj}-2A_3^{kij}+A_2^{ikj}+A_2^{kij})-133(A_4^{ijk}+A_4^{ikj})\nn\\
g_3^{ijk}&=&-6(A_1^{ijk}+A_1^{jik}-2A_3^{ikj}-2A_3^{kij}+A_2^{ikj}+A_2^{kij})+15(A_4^{ijk}+A_4^{ikj})\nn\\
g_{4,1}^{ijk}&=&-11 (A_1^{ijk}+A_1^{jik}+7 A_2^{ikj}+7 A_2^{kij})-133 (A_4^{ijk}+A_4^{ikj}) \nn\\
g_{4,2}^{ijk}&=&-84 (A_3^{ijk}+A_3^{jik})-21 A_5^{ijk}-6 A_6^{ijk} + (j\leftrightarrow k)\nn\\
g_{4,3}^{ijk}&=&36 (A_1^{ijk}+A_1^{jik})-6 A_5^{ijk}-21 A_6^{ijk} + (j\leftrightarrow k)\nn\\
g_{4,4}^{ijk}&=&-11 (A_1^{ikj}+A_1^{kij}+7 A_2^{ijk}+7 A_2^{jik})-133 (A_4^{ijk}+A_4^{ikj}) \nn\\
g_{5,1}^{ijk}&=&-15 (A_1^{ijk}+A_1^{jik}+7 A_2^{ikj}+7 A_2^{kij})+15 (A_4^{ijk}+A_4^{ikj}) \nn\\
g_{5,2}^{ijk}&=&-4 (A_3^{ijk}+A_3^{jik}-18 A_2^{ijk}-18 A_2^{jik})+71 A_5^{ijk}-62 A_6^{ijk} + (j\leftrightarrow k)\nn\\
g_{5,3}^{ijk}&=&12 (A_1^{ijk}+A_1^{jik})-62 A_5^{ijk}+71 A_6^{ijk} + (j\leftrightarrow k)\nn\\
g_{5,4}^{ijk}&=&-15 (A_1^{ikj}+A_1^{kij}+7 A_2^{ijk}+7 A_2^{jik})+15 (A_4^{ijk}+A_4^{ikj})\nn
}
}
{
\eqa{
h_1^{ijk}&=&-85 (B_1^{ijk}+B_1^{jik})+11(B_2^{ijk}+B_2^{jik})\ ;\quad
h_2^{ijk}=-68 (B_1^{ijk}+B_1^{jik})\ ;\nn\\
h_3^{ijk}&=&12 (B_1^{ijk}+B_1^{jik})\ ;\quad
h_{4,1}^{ijk}=-84 (B_1^{ijk}+B_1^{jik})\ ;\quad
h_{4,2}^{ijk}=-12 (B_2^{ijk}+B_2^{jik})\ ;\nn\\
h_{4,3}^{ijk}&=&77 (B_1^{ijk}+B_1^{jik})-11 (B_2^{ijk}+B_2^{jik})\ ;\quad
h_{5,1}^{ijk}=-4 (B_1^{ijk}+B_1^{jik})\ ;\nn\\
h_{5,2}^{ijk}&=&-144 (B_1^{ijk}+B_1^{jik})+20 (B_2^{ijk}+B_2^{jik})\ ;\quad
h_{5,3}^{ijk}=105 (B_1^{ijk}+B_1^{jik})-15 (B_2^{ijk}+B_2^{jik})\ ;\nn
}
}
{
\eqa{
m_1^{ijk}&=&-2 F^k (C_1^{ij}+C_1^{ji})+11 F^k(C_2^{ij}+C_2^{ji})\ ;\quad
m_2^{ijk}=-17 F^k (C_1^{ij}+C_1^{ji})\ ;\nn\\
m_3^{ijk}&=&3 F^k (C_1^{ij}+C_1^{ji})\ ;\quad
m_{4,2}^{ijk}=-11 F^k (C_2^{ij}+C_2^{ji})\ ;\quad
m_{5,2}^{ijk}=-15 F^k (C_2^{ij}+C_2^{ji})\ ;\nn\\
m_{4,1}^{ijk}&=&-21 F^k (C_1^{ij}+C_1^{ji})-6 F^k(C_2^{ij}+C_2^{ji})\ ;\quad
m_{5,1}^{ijk}=-F^k (C_1^{ij}+C_1^{ji})+10 F^k(C_2^{ij}+C_2^{ji})\ ;\nn
}
}
{
\eqa{
n_1^{ijkl}&=&-2 (D_1^{ijkl}+D_1^{jikl})+11 (D_2^{ijkl}+D_2^{jikl})\ ;\quad
n_2^{ijkl}=-17 (D_1^{ijkl}+D_1^{jikl})\ ;\nn\\
n_3^{ijkl}&=&3 (D_1^{ijkl}+D_1^{jikl})\ ;\quad
n_{4,2}^{ijkl}=-11 (D_2^{ijkl}+D_2^{jikl})\ ;\quad
n_{5,2}^{ijkl}=-15 (D_2^{ijkl}+D_2^{jikl})\ ;\nn\\
n_{4,1}^{ijkl}&=&-21 (D_1^{ijkl}+D_1^{jikl})-6 (D_2^{ijkl}+D_2^{jikl})\ ;\quad
n_{5,1}^{ijkl}= -(D_1^{ijkl}+D_1^{jikl})+10(D_2^{ijkl}+D_2^{jikl})\ ;\nn
}
}
{
\eqa{
p_1^{ijkl}&=&  4 E^{il} E^{jk}  \ ;\quad
p_{4,1}^{ijkl}=   -2 E^{ij} E^{kl}\ ;\quad
p_{4,2}^{ijkl}=   342 E^{ik} E^{jl}\ ;\nn\\
p_2^{ijk}&=&  -8 E^{il} E^{jk} +171 E^{ik} E^{jl} -E^{ij} E^{kl}  \ ;\quad
p_{5,1}^{ijkl}=   6 E^{ij} E^{kl}\ ;\nn\\
p_3^{ijk}&=&  -8 E^{il} E^{jk} - E^{ik} E^{jl} +3 E^{ij} E^{kl}  \ ;\quad
p_{5,2}^{ijkl}=  -2 E^{ik} E^{jl}\ ;\nn\\
&&\nn
}
}
with the functions $A_i^{ijk}, B_i^{ijk}, C_i^{ij}, D_i^{ijkl}, E_i^{ij}$ and $F^{k}$ given by

{
\eqa{
A_1^{ijk}&=&\frac{1}{2(x_i-1)^2(x_j-1)^2(x_k-1)(x_i-x_j)}\,\bigg [ 8(x_j-1)^2(x_i-x_k)^2\Li(1-{\textstyle \frac{x_i}{x_k}})\nn\\
&&-8(x_i-1)^2(x_j-1)^2\Li(1-x_i)-8 (x_j-1)^2(x_k-1) x_i\log{x_i}\nn\\
&&+(x_k-1)(x_i-x_j)(2x_ix_j-x_ix_k-x_jx_k-x_i-x_j+2x_k)(4\Li(1-x_k)-\log^2 x_k )\nn\\
&&+6(x_j-1)^2(x_i-x_k)^2 \log^2{x_k} -4(x_i-1)(x_j-1)(x_k-1)(x_i-x_j)\nn\\
&&-4(x_i-1)(x_j-1)(x_i-x_j)x_k\log{x_k} -8(x_j-1)^2 x_i^2\log{x_i}\log{x_k}\bigg]\nn\\[2mm]
A_2^{ijk}&=&-\frac{x_i}{7} A_1^{ijk}+\frac{1}{14(x_i-1)^2(x_j-1)^2(x_k-1)(x_i-x_j)}\,\bigg[ -8 (x_j-1)^2(x_k-1)x_i^2\log^2{x_i}\nn\\
&&+(x_i-1)^2(x_i-x_j)(x_k-1)(2x_j-x_k-1) (4\Li(1-x_k)-\log^2{x_k})\nn\\
&&-4(x_j-1)(x_i-x_j)(x_i^2-1)x_k \log{x_k}-8(x_j-1)^2(2x_k\log{x_k}-5x_k+5)x_i^2\log{x_i}\nn\\
&&-4(x_i-1)(x_i-6)(x_j-1)(x_k-1)(x_i-x_j)\bigg] \nn\\[2mm]
A_3^{ijk}&=&\frac{1}{2} A_1^{ijk}+\frac{1}{2} A_2^{ijk}+\frac{2}{(x_i-1)^2(x_j-1)(x_k-1)(x_i-x_j)}\,\bigg[ (x_j-1)(x_k-1)x_i\log^2{x_i}\nn\\
&&+2(x_j-1)(x_k\log{x_k}-2x_k+2)x_i\log{x_i}+(x_i-1)(x_i-x_j)(x_k\log{x_k}-2x_k+2)\bigg]\nn\\[2mm]
A_4^{ijk}&=&\frac{1}{(x_i-1)^2(x_j-1)(x_k-1)(x_j-x_k)}\,\bigg [ 8(x_k-1)(x_i-x_j)^2 \Li(1-{\textstyle \frac{x_i}{x_j}})\nn\\
&&+4(x_i-1)^2(x_j-x_k)\Li(1-x_i)-8(x_j-1)(x_k-1)(2 x_i-x_j-1)\Li(1-x_j)\nn\\
&&-(x_j x_i^2+x_k x_i^2-2 x_i^2+2 x_i x_j+2 x_i x_k-4 x_i x_j x_k-x_j^2+x_j x_k^2-x_k^2+x_j^2 x_k) \log^2{x_i}\nn\\
&&+4(x_k-1)(x_i-x_j)^2\log^2{x_j}-8(x_k-1)x_i^2\log{x_i}\log{x_j}\nn\\
&&+8(x_i-1)(x_k-1)x_j\log{x_j}\bigg]\nn\\[2mm]
A_5^{ijk}&=&-\frac{1}{2}A_4^{ijk}+\frac{8 (x_i-1-x_i\log{x_i})x_j\log{x_j}}{(x_i-1)^2(x_j-1)(x_j-x_k)}\nn\\[2mm]
A_6^{ijk}&=&-\frac{x_j}{2}A_4^{ijk}+\frac{4(x_i-1)^2\Li(1-x_i)-(x_i-x_k)^2\log^2{x_i}+4(x_k-1)(x_i\log{x_i}-x_i+1)}{2(x_i-1)^2(x_k-1)}\nn\\[2mm]
B_1^{ijk}&=&\frac{3}{(x_i-1)^3(x_i-x_j)}\,\bigg[
4 (x_i - x_k)^2 \Li(1-{\textstyle \frac{x_i}{x_k}}) + x_i (-4 (2 + (-3 + x_i) x_i) x_k - x_i \log{x_i}^2 \nn\\
&&- 4 \log{x_i} (x_k + x_i \log{x_k}) + 2 \log{x_k} (-2 (2 + (-3 + x_i) x_i) x_k + (x_i - 2 x_k\nn\\
&&+ (3 + (-3 + x_i) x_i) x_k^2) \log{x_k}) + 4 (x_i - 2 x_k + (3 + (-3 + x_i) x_i) x_k^2) \Li(1-x_k))
\bigg]\nn\\[2mm]
B_2^{ijk}&=&-(x_i-6) B_1^{ijk}+\frac{6 x_i x_k (x_k \log^2{x_k}-2 \log{x_k}+2x_k\Li(1-x_k)-2)}{x_i-x_j}\nn\\[2mm]
C_1^{ij}&=&32\ \frac{2x_i \log{x_i} - x_i^2+1}{(x_i-1)^2(x_i-x_j)}\quad ;\qquad
C_2^{ij}=-16\ \frac{2x_i^2 \log{x_i} - 3 x_i^2+4x_i-1}{(x_i-1)^2(x_i-x_j)}\nn\\[2mm]
D_1^{ijkl}&=&\frac{64 x_l (\log(m_{\tilde g}^2/\mu^2)+\log{x_l}-1)}{(x_i-1)^2(x_k-1)^3(x_i-x_j)(x_k-x_i)}\,\bigg[
2 x_i (x_k-1)^3 \log{x_i} \nn\\
&&+ (x_i-1) (( x_k-1)(-x_k ( x_k-3) + x_i^2 (x_k+1) - x_i (3 + x_k^2)) - 2 ( x_i-1)^2 x_k \log{x_k})
\bigg]\nn\\[2mm]
D_2^{ijkl}&=&\frac{32 x_l (\log(m_{\tilde g}^2/\mu^2)+\log{x_l}-1)}{(x_i-1)^2(x_k-1)^3(x_i-x_j)(x_k-x_i)}\,\bigg[
-2 x_i^2( x_k-1)^3 \log{x_i} \nn\\
&&+ ( x_i-1)(( x_k-1)(x_i + x_i^2(1 - 3 x_k) + 3 x_i x_k^2 - x_k ( x_k+1)) + 2 (x_i-1)^2 x_k^2 \log{x_k})
\bigg]\nn\\[2mm]
E^{ij}&=&\frac{2 x_i\log{x_i}}{(x_i-1)(x_i-x_j)}-\frac{2 x_j\log{x_j}}{(x_j-1)(x_i-x_j)}\quad ;\qquad
F^k=\frac{x_k^2-4 x_k+3-2x_k(x_k-2)\log{x_k}}{4 (x_k-1)^2}\nn
}
}

Finally, the functions $f^{ij}$ appearing in the part with four squark rotation matrices are given by:

{
\eqa{
f_1^{ij}&=&\frac{-1}{(x_i-1)^3(x_j-1)^3(x_i-x_j)^2}\,\bigg\{
\Big[ 8 (x_i - x_j)  (x_j-1)^3 (-115 - 72 x_i + 399 x_i^2 - 212 x_i^3\nn\\
&& + (-32 - 220 x_i + 74 x_i^2 + 88 x_i^3) \log{x_i})\Big] \log(m_{\tilde g}^2/\mu^2)\nn\\
&&+\Big[
4 (x_i - x_j) x_j (2395 + x_j (-3587 + 3579 x_i -  (x_i-1) (1189 + 11 x_i) x_j + 3  (x_i-1)^2 x_j^2))
\Big]\nn\\
&&+\Big[
 (x_j-1) (-3 x_i^6 + x_i^5 (53 + 34 x_j) + x_i^4 (700 - 1584 x_j + 773 x_j^2) \nn\\
&&+ x_i^3 (102 - 1438 x_j + 2107 x_j^2 - 1137 x_j^3)+ x_i^2 (-72 + 392 x_j + 422 x_j^2 + 231 x_j^3 - 34 x_j^4) \nn\\
&& + x_j^2 (120 + 40 x_j + 50 x_j^2 - 3 x_j^3)+ x_i x_j (256 - 1082 x_j + 86 x_j^2 - 16 x_j^3 + 3 x_j^4))
\Big]\log^2 x_i\nn\\
&&-2 \Big[
2  (x_i-1) (x_i - x_j)  (x_j-1) (-983 + x_i (1232 + x_i (419 + x_i (-50 + 3 x_i))) + 1890 x_j \nn\\
&&- x_i (2263 + x_i (838 + 31 x_i)) x_j + (-1114 + 5 x_i (274 + 73 x_i)) x_j^2)
\Big]\Li(1-x_i)\nn\\
&&+\Big[
2  (x_i-1) (x_i - x_j)  (x_j-1) (-3 x_i^4 + x_i^3 (50 + 31 x_j) + x_j (120+ x_j (40 + (50 - 3 x_j) x_j)) \nn\\
&& + x_i^2 (40 - 80 x_j + 94 x_j^2) + x_i (120 + x_j (-410 + x_j (-80 + 31 x_j))))
\Big]\Li(1-{\textstyle \frac{x_j}{x_i}})\nn\\
&&+\Big[
-4 (x_i - x_j)  (x_j-1)^2 (-64  (x_j-1) + x_i (1563 + x_i (-311 + x_i (-119 + 3 x_i)) \nn\\
&&- 1575 x_j+ x_i (341 + 98 x_i) x_j + 3  (x_i-1) x_j^2))
\Big]\log x_i\nn\\
&&+\Big[
 (x_i-1)  (x_j-1) (-3 x_i^5 + x_i^4 (50 + 34 x_j) + x_i^3 (40 + x_j (-98 + 151 x_j))\nn\\
&& + x_j^2 (120 + x_j (40 + (50 - 3 x_j) x_j)) +  2 x_i x_j (32 + x_j (-81 + x_j (-49 + 17 x_j)))\nn\\
&&+ x_i^2 (120 + x_j (-162 + x_j (-328 + 151 x_j))))
\Big]\log x_i \log x_j\bigg\}\nn\\[2mm]
f_2^{ij}&=&\frac{-1}{(x_i-1)^3(x_j-1)^3(x_i-x_j)^2}\,\bigg\{
\Big[
8 (x_i - x_j)  (x_j-1)^3 (-515 + 889 x_i - 437 x_i^2 + 63 x_i^3 \nn\\
&&+ 2 (-136 - 5 x_i + 39 x_i^2) \log{x_i})\Big] \log(m_{\tilde g}^2/\mu^2)\nn\\
&&+\Big[
4 (x_i - x_j) x_j (7419 + x_j (-12118 + 14097 x_i +  (x_i-1) (-4699 + 1979 x_i) x_j))
\Big]\nn\\
&&+\Big[
2  (x_j-1) (-261 x_i^5 + x_i^4 (547 + 668 x_j) + x_i x_j (1088 - 3764 x_j + 1470 x_j^2 + 261 x_j^3) \nn\\
&&+ 2 x_i^3 (909 - 3013 x_j + 979 x_j^2)+ x_j^2 (146 + 286 x_j - 261 x_j^2)  \nn\\
&&- 2 x_i^2 (610 - 809 x_j - 2156 x_j^2 + 1320 x_j^3))
\Big]\log^2 x_i\nn\\
&&-2 \Big[
4  (x_i-1) (x_i - x_j)  (x_j-1) (-918 + 261 x_i^3 + (1429 - 682 x_j) x_j \nn\\
&&- 11 x_i^2 (26 + 37 x_j) + x_i (1325 + 38 x_j (-47 + 28 x_j)))
\Big]\Li(1-x_i)\nn\\
&&+\Big[
-4  (x_i-1) (x_i - x_j)  (x_j-1) (261 x_i^3 - 11 x_i^2 (26 + 37 x_j) \nn\\
&&+ x_i (-146 + (1156 - 407 x_j) x_j) + x_j (-146 + x_j (-286 + 261 x_j)))
\Big]\Li(1-{\textstyle \frac{x_j}{x_i}})\nn\\
&&+\Big[
4 (x_i - x_j)  (x_j-1)^2 (544  (x_j-1)\nn\\ 
&&+ x_i (-3823 + x_i (1129 + 518 x_i - 2165 x_j) + 4341 x_j))
\Big]\log x_i\nn\\
&&+\Big[
-2  (x_i-1)  (x_j-1) (261 x_i^4 + x_i^3 (-286 + 372 x_j) + 2 x_i x_j (139 + 6 x_j (25 + 31 x_j)) \nn\\
&&- 2 x_i^2 (73 + 10 x_j (-15 + 64 x_j)) + x_j^2 (-146 + x_j (-286 + 261 x_j)))
\Big]\log x_i \log x_j \bigg\}\nn\\[2mm]
f_3^{ij}&=&\frac{-1}{(x_i-1)^3(x_j-1)^3(x_i-x_j)^2}\,\bigg\{
\Big[
8 (x_i - x_j)  (x_j - 1)^3 (121 - 187 x_i + 47 x_i^2 + 19 x_i^3 \nn\\
&&+ (48 + 62 x_i - 74 x_i^2) \log{x_i})
\Big] \log(m_{\tilde g}^2/\mu^2)\nn\\
&&+\Big[
60 (x_i - x_j) x_j (-47 + x_j (62 - 45 x_i +  (x_i - 1) (15 + 17 x_i) x_j))
\Big]\nn\\
&&+\Big[
-2  (x_j - 1) (9 x_i^5 + x_i^4 (-95 + 68 x_j) + x_j^2 (86 - 86 x_j + 9 x_j^2)  \nn\\
&&+ 2 x_i^3 (143 - 191 x_j + 57 x_j^2) + x_i x_j (192 - 556 x_j + 346 x_j^2 - 9 x_j^3) \nn\\
&& - 2 x_i^2 (22 + 173 x_j - 412 x_j^2 + 208 x_j^3))
\Big]\log^2 x_i\nn\\
&&-2 \Big[
4  (x_i - 1) (x_i - x_j)  (x_j - 1) (162 + 9 x_i^3 + x_i^2 (-86 + 77 x_j) + x_j (-247 + 94 x_j)  \nn\\
&&+ x_i (-239 + 478 x_j - 248 x_j^2))
\Big]\Li(1-x_i)\nn\\
&&+\Big[
-4  (x_i - 1) (x_i - x_j)  (x_j - 1) (x_i + x_j) (86 + 9 x_i^2 + x_j (-86 + 9 x_j)  \nn\\
&&+ x_i (-86 + 68 x_j))
\Big]\Li(1-{\textstyle \frac{x_j}{x_i}})\nn\\
&&+\Big[
4 (x_i (557 + x_i (-395 + 222 x_i - 49 x_j) - 335 x_j) - 96  (x_j - 1)) (x_i - x_j)  (x_j - 1)^2 \Big]\log x_i\nn\\
&&+\Big[
-2  (x_i - 1)  (x_j - 1) (9 x_i^4 + x_i^3 (-86 + 308 x_j) - 2 x_i^2 (-43 + 74 x_j + 200 x_j^2)  \nn\\
&&+ x_j^2 (86 + x_j (-86 + 9 x_j)) + 2 x_i x_j (31 + 2 x_j (-37 + 77 x_j)))
\Big]\log x_i \log x_j
\bigg\}\nn\\[2mm]
f_{4,1}^{ij}&=&\frac{-1}{(x_i-1)^3(x_j-1)^3(x_i-x_j)^2}\,\bigg\{
\Big[
-48 (x_i - x_j) (x_j - 1)^3 (167 - 333 x_i + 237 x_i^2 - 71 x_i^3  \nn\\
&&+ (112 - 91 x_i + 50 x_i^2 + x_i^3) \log{x_i})
\Big] \log(m_{\tilde g}^2/\mu^2)\nn\\
&&+\Big[
48 (x_i - x_j) x_j (1354 + x_j (-2225 + 2616 x_i + 4 (x_i - 1) (-217 + 97 x_i) x_j + 3 (x_i - 1)^2 x_j^2))
\Big]\nn\\
&&+\Big[
-6 (x_j - 1) (6 x_i^6 + x_i^5 (9 - 30 x_j) + x_i^4 (121 - 254 x_j + 184 x_j^2) + 3 x_j^2 (2 + 5 x_j^2 + 2 x_j^3)  \nn\\
&&- 2 x_i^3 (719 - 1324 x_j + 530 x_j^2 + 132 x_j^3) - x_i x_j (896 - 1750 x_j + 908 x_j^2 + 45 x_j^3 + 6 x_j^4) \nn\\
&&+ 2 x_i^2 (339 + 202 x_j - 1376 x_j^2 + 898 x_j^3 + 15 x_j^4))
\Big]\log^2 x_i\nn\\
&&-2 \Big[
12 (x_i - 1) (x_i - x_j) (x_j - 1) (-620 + 6 x_i^4 + 3 x_i^3 (5 - 8 x_j) + (1219 - 572 x_j) x_j  \nn\\
&&+ x_i^2 (-50 + (103 - 26 x_j) x_j) + x_i (613 + 2 x_j (-613 + 281 x_j)))
\Big]\Li(1-x_i)\nn\\
&&+\Big[
-36 (x_i - 1) (x_i - x_j) (x_j - 1) (2 x_i^4 + x_i^3 (5 - 8 x_j) + 2 x_j + x_j^3 (5 + 2 x_j)  \nn\\
&&+ x_i^2 x_j (1 + 8 x_j) + x_i (2 + x_j (-12 + x_j - 8 x_j^2)))
\Big]\Li(1-{\textstyle \frac{x_j}{x_i}})\nn\\
&&+\Big[
-24 (x_i - x_j) (x_j - 1)^2 (-224 (x_j - 1) + x_i (1305 + x_i (-610 + x_i (35 + 6 x_i)) \nn\\
&& - 1248 x_j + 4 x_i (127 + x_i) x_j + 6 (x_i - 1) x_j^2))
\Big]\log x_i\nn\\
&&+\Big[
-6 (x_i - 1) (x_j - 1) (6 x_i^5 + 15 x_i^4 (1 - 2 x_j) + 4 x_i^3 x_j (-31 + 16 x_j)  \nn\\
&&+ 3 x_j^2 (2 + x_j^2 (5 + 2 x_j)) - 2 x_i x_j (112 + x_j (-126 + x_j (62 + 15 x_j))) \nn\\
&&+ 2 x_i^2 (3 + x_j (126 + x_j (-77 + 32 x_j))))
\Big]\log x_i \log x_j
\bigg\}\nn\\[2mm]
f_{4,2}^{ij}&=&\frac{-1}{(x_i-1)^3(x_j-1)^3(x_i-x_j)^2}\,\bigg\{
\Big[4 (x_i - x_j)  (x_j - 1)^3 (597 - 1533 x_i + 1539 x_i^2 - 603 x_i^3 \nn\\
&& + 2 x_i (352 - 509 x_i + 289 x_i^2) \log{x_i})\Big] \log(m_{\tilde g}^2/\mu^2)\nn\\
&&+\Big[-32 (x_i - x_j) x_j (487 + x_j (-754 + 801 x_i +  (x_i - 1) (-267 + 47 x_i) x_j))\Big]\nn\\
&&+\Big[2  (x_j - 1) (27 x_j^3 (7 + 9 x_j) - 27 x_i^5 (-9 + 25 x_j) - 27 x_i x_j^2 (5 + 41 x_j + 34 x_j^2) \nn\\
&& + x_i^4 (-880 + 1409 x_j + 1631 x_j^2) + x_i^3 (65 + 1697 x_j - 5047 x_j^2 - 1035 x_j^3)  \nn\\
&&+ 5 x_i^2 x_j (-143 + 367 x_j + 505 x_j^2 + 135 x_j^3))\Big]\log^2 x_i\nn\\
&&-2 \Big[4  (x_i - 1) (x_i - x_j)  (x_j - 1) (709 + 27 x_i^3 (-9 + 25 x_j) + x_j (-1175 + 34 x_j)  \nn\\
&&+ x_i^2 (871 - x_j (1445 + 722 x_j)) + x_i (-1067 + x_j (1405 + 958 x_j)))\Big]\Li(1-x_i)\nn\\
&&+\Big[-108  (x_i - 1) (x_i - x_j)  (x_j - 1) (-(x_j^2 (7 + 9 x_j)) + x_i^3 (-9 + 25 x_j)  \nn\\
&&+ x_i^2 (-7 + (25 - 66 x_j) x_j) + x_i x_j (-2 + 25 x_j (1 + x_j)))\Big]\Li(1-{\textstyle \frac{x_j}{x_i}})\nn\\
&&+\Big[8 x_i (x_i - x_j)  (x_j - 1)^2 (1061 - 453 x_j + x_i (-455 - 761 x_j + x_i (274 + 334 x_j)))\Big]\log x_i\nn\\
&&+\Big[2  (x_i - 1)  (x_j - 1) (27 x_i^4 (9 - 25 x_j) + 27 x_j^3 (7 + 9 x_j) - x_i x_j^2 (487 + 27 x_j (34 + 25 x_j)) \nn\\
&& + x_i^3 (189 + x_j (-918 + 377 x_j)) + x_i^2 x_j (-487 + x_j (2542 + 377 x_j)))\Big]\log x_i \log x_j
\bigg\}\nn\\[2mm]
f_{5,1}^{ij}&=&\frac{-1}{(x_i-1)^3(x_j-1)^3(x_i-x_j)^2}\,\bigg\{
\Big[16 (x_i - x_j)  (x_j - 1)^3 (-77 - 81 x_i + 321 x_i^2 - 163 x_i^3  \nn\\
&&+ (-16 - 179 x_i + 58 x_i^2 + 65 x_i^3) \log{x_i})\Big] \log(m_{\tilde g}^2/\mu^2)\nn\\
&&+\Big[16 (x_i - x_j) x_j (1258 + x_j (-2015 + 2292 x_i + 32  (x_i - 1) (-23 + 8 x_i) x_j + 21  (x_i - 1)^2 x_j^2))\Big]\nn\\
&&+\Big[-2  (x_j - 1) (42 x_i^6 - x_i^5 (257 + 178 x_j) + x_i^4 (-689 + 2414 x_j - 792 x_j^2)  \nn\\
&&+ x_j^2 (-54 - 160 x_j - 215 x_j^2 + 42 x_j^3) + 2 x_i^3 (119 + 292 x_j - 1182 x_j^2 + 660 x_j^3)  \nn\\
&&+ x_i x_j (-128 + 634 x_j + 780 x_j^2 + 37 x_j^3 - 42 x_j^4)  \nn\\
&&+ 2 x_i^2 (21 - 410 x_j + 352 x_j^2 - 658 x_j^3 + 89 x_j^4))\Big]\log^2 x_i\nn\\
&&-2 \Big[4  (x_i - 1) (x_i - x_j)  (x_j - 1) (-756 + x_i (787 + x_i (178 + x_i (-215 + 42 x_i))) \nn\\
&& + 1589 x_j - x_i (1382 + x_i (143 + 136 x_i)) x_j + 2 (-610 + x_i (551 + 77 x_i)) x_j^2)\Big]\Li(1-x_i)\nn\\
&&+\Big[4  (x_i - 1) (x_i - x_j)  (x_j - 1) (-42 x_i^4 + x_i^3 (215 + 136 x_j) + x_i^2 (160 + x_j (-533 + 184 x_j))  \nn\\
&&+ x_j (54 + x_j (160 + (215 - 42 x_j) x_j)) \nn\\
&&+ x_i (54 + x_j (-164 + x_j (-533 + 136 x_j))))\Big]\Li(1-{\textstyle \frac{x_j}{x_i}})\nn\\
&&+\Big[-8 (x_i - x_j)  (x_j - 1)^2 (-32  (x_j - 1) + x_i (1551 + 7 x_i (-82 + x_i (-13 + 6 x_i))  \nn\\
&&- 1728 x_j + 4 (253 - 53 x_i) x_i x_j + 42  (x_i - 1) x_j^2))\Big]\log x_i\nn\\
&&+\Big[2  (x_i - 1)  (x_j - 1) (-42 x_i^5 + x_i^4 (215 + 178 x_j) + 4 x_i^3 (40 + x_j (-183 + 32 x_j))  \nn\\
&&+ x_j^2 (54 + x_j (160 + (215 - 42 x_j) x_j)) + 2 x_i^2 (27 + x_j (-18 + x_j (59 + 64 x_j)))  \nn\\
&&+ 2 x_i x_j (16 + x_j (-18 + x_j (-366 + 89 x_j))))\Big]\log x_i \log x_j
\bigg\}\nn\\[2mm]
f_{5,2}^{ij}&=&\frac{-1}{(x_i-1)^3(x_j-1)^3(x_i-x_j)^2}\,\bigg\{
\Big[-60 (x_i - x_j)  (x_j - 1)^3 (-3 (5 + 19 x_i - 45 x_i^2 + 21 x_i^3)  \nn\\
&&+ 2 x_i (-32 + 7 x_i + 13 x_i^2) \log{x_i})\Big] \log(m_{\tilde g}^2/\mu^2)\nn\\
&&+\Big[96 (x_i - x_j) x_j (-157 + x_j (214 - 171 x_i +  (x_i - 1) (57 + 43 x_i) x_j))\Big]\nn\\
&&+\Big[-30  (x_j - 1)^2 (9 x_i^5 - 9 x_j^3 + 9 x_i x_j^2 (3 + 2 x_j) + 5 x_i^4 (-16 + 7 x_j)  \nn\\
&&+ x_i^3 (19 + 134 x_j - 63 x_j^2) - x_i^2 x_j (65 + 16 x_j + 9 x_j^2))\Big]\log^2 x_i\nn\\
&&-2 \Big[60  (x_i - 1) (x_i - x_j)  (x_j - 1)^2 (-71 + x_i (97 + x_i (19 + 9 x_i - 46 x_j)\nn\\
&&- 70 x_j) + 62 x_j)\Big]\Li(1-x_i)\nn\\
&&+\Big[-540  (x_i - 1)^2 (x_i - x_j)^3  (x_j - 1)^2\Big]\Li(1-{\textstyle \frac{x_j}{x_i}})\nn\\
&&+\Big[24 x_i (x_i - x_j)  (x_j - 1)^2 (515 - 435 x_j + x_i (-137 - 23 x_j + x_i (22 + 58 x_j)))\Big]\log x_i\nn\\
&&+\Big[-30  (x_i - 1)  (x_j - 1) (9 x_i^4  (x_j - 1) - 9  (x_j - 1) x_j^3 + x_i^3 (3 + x_j) (3 + 5 x_j)  \nn\\
&&+ x_i x_j^2 (5 + 9 x_j (2 + x_j)) + x_i^2 x_j (5 + x_j (-74 + 5 x_j)))\Big]\log x_i \log x_j
\bigg\}\nn
}
}

\section{Flavor changing quark self-energies: Relationship between the tree-level and on-shell definitions of the super-CKM basis}
\label{app3}

In this appendix we elaborate on the relationship between the two different definitions of the super-CKM basis addressed in Section \ref{sec:ren}.

The flavor changing self-energies of Fig.~\ref{FCSE} can be written as\footnote{These corrections can be found in Ref.~\cite{arXiv:0810.1613}. A discussion of how these can be absorbed into wave-function counterterms is given in Ref.~\cite{0907.5408}.}
\eq{\Sigma_{q'q}(p)=\Sigma_{q'q}^{RL}(p^2)P_L+\Sigma_{q'q}^{LR}(p^2)P_R+/\!\!\!p\,[\,\Sigma_{q'q}^{LL}(p^2)P_L+\Sigma_{q'q}^{RR}(p^2)P_R\,].}
When inserted as external legs, the quark propagator will provide a chiral enhancement with the chirality-flipping part of the self-energy, and we will keep only this contribution. Moreover we will expand in the external momentum, which is justified by the fact that the SUSY masses are much higher than the external quark masses. Then, the self energies we are interested in are given by
\eq{\Sigma_{q'q}^{RL, LR}(0)=\frac{2\alpha_s}{3\pi}m_{\tilde g} \sum_k g(\tilde{m}_k^2/m_{\tilde g}^2)\ \Gamma_{R,L}^{kq'*}\,\Gamma_{L,R}^{kq}}
with $\ g(x)=(1-x+x\log x)/(1-x)$.

For the case of $B_s$ mixing, the corrected couplings of the external quarks are,\\

\begin{center}
\begin{minipage}{5cm}
\psfrag{b}{$b_\alpha$} \psfrag{d}{$d,s$}\psfrag{a}{$a$}\psfrag{i}{$i_\beta$}
\includegraphics{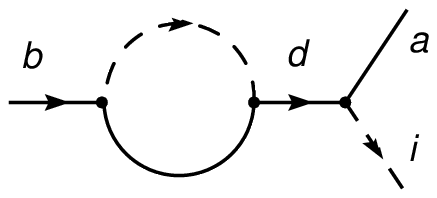}
\end{minipage}
$\displaystyle =\ -ig_s\sqrt{2}\ T^a_{\beta\alpha}\ [\,\Pi_L^{ib} P_L-\Pi_R^{ib} P_R\,]\ u_b^\alpha$
\end{center}

\begin{center}
\begin{minipage}{5cm}
\psfrag{d}{$s_\alpha$} \psfrag{b}{$d,b$}\psfrag{a}{$a$}\psfrag{i}{$i_\beta$}
\includegraphics{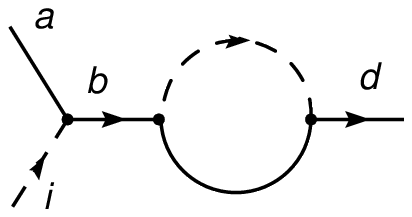}
\end{minipage}
$\displaystyle =\ -ig_s\sqrt{2}\ T^a_{\beta\alpha}\ \bar{u}_s^\alpha\ [\,\Pi_L^{is*} P_R-\Pi_R^{is*} P_L\,]$\\[5mm]
\end{center}
where, in the approximation $m_b\gg m_s\gg m_d$, and keeping only the chirally enhanced terms,
\eqa{
\Pi_L^{ib}&=&\Gamma_L^{id}\ \Sigma_{db}^{LR}/m_b +\Gamma_L^{is}\ \Sigma_{sb}^{LR}/m_b\\[2mm]
\Pi_L^{is*}&=&\Gamma_L^{id*}\ \Sigma_{sd}^{RL}/m_s -\Gamma_L^{ib*}\ \Sigma_{sb}^{LR}/m_b
}
and correspondingly for $L\leftrightarrow R$.

At this point, we denote by $\Gamma^{(0)}$ the squark rotation matrices associated with the tree-level definition of the super-CKM basis, and $\Gamma$ the rotation matrices in the on-shell case. At tree-level both coincide; at the one loop level, the relationship between both definitions is given by
\eqa{
\Gamma_L^{ib}&=&\Gamma_L^{(0)\,ib} +\Gamma_L^{id}\ \Sigma_{db}^{LR}/m_b +\Gamma_L^{is}\ \Sigma_{sb}^{LR}/m_b\nn\\[2mm]
\Gamma_L^{is*}&=&\Gamma_L^{(0)\,is*} +\Gamma_L^{id*}\ \Sigma_{sd}^{RL}/m_s -\Gamma_L^{ib*}\ \Sigma_{sb}^{LR}/m_b\label{rel1}
}
and correspondingly for $L\leftrightarrow R$. The full results for the NLO Wilson coefficients in terms of $\Gamma^{(0)}$ --corresponding to the tree-level definition of the super-CKM basis-- can be obtained by making the substitutions of Eq.~(\ref{rel1}) in the LO expressions given in Eq.~(\ref{LOWC}).

In the degenerate MIA case, these corrections cancel. Consider for example the coefficient $C_1$ in Eq.~(\ref{LOWC}). Making the substitutions of Eq.~(\ref{rel1}) in the rotation matrices with index $i$, and making the reduction to the MIA (see Section~\ref{sec:MIA1}) gives, at order $\alpha_s^3$, two terms:
\eq{\sum_i f(x_i,\cdots)\Pi_L^{is*}\,\Gamma_L^{ib}= -\frac{2\alpha_s}{3\pi}f(M_s^2/m_{\tilde g}^2,\cdots)\,g'(M_s^2/m_{\tilde g}^2)\,\frac{M_s^2}{m_{\tilde g} m_b}\,\delta^{LR}_{sb}}
\eq{\sum_i f(x_i,\cdots)\Gamma_L^{is*}\,\Pi_L^{ib}= \frac{2\alpha_s}{3\pi}f(M_s^2/m_{\tilde g}^2,\cdots)\,g'(M_s^2/m_{\tilde g}^2)\,\frac{M_s^2}{m_{\tilde g} m_b}\,\delta^{LR}_{sb}}
and both cancel due to the relative signs in Eqs.~(\ref{rel1}). The same happens with the other two terms arising from the $\Gamma_L^{jq}$, and in the other Wilson coefficients. Therefore, the two bases coincide in the mass insertion approximation with degenerate squarks.

Finally, the relationship between the mass insertions defined in each basis can be deduced from Eqs.~(\ref{rel1}) and (\ref{rotrel}). As we have just seen, both are equivalent in the degenerate case. In the non-denerate case, one such relation is (at leading order in the mass insertion expansion),
\eq{\Delta^{LL}_{sb}=\Delta^{(0)\,LL}_{sb}-\frac{2\alpha_s}{3\pi} \frac{m_{\tilde g}}{m_b}\ \frac{g(X_{\tilde{s}_L}/m_{\tilde g}^2)-g(X_{\tilde{b}_R}/m_{\tilde g}^2)}{X_{\tilde{s}_L}-X_{\tilde{b}_R}} \ [X_{\tilde{b}_L}-X_{\tilde{s}_L}] \ \Delta_{sb}^{LR}\ ,
\label{relDelts}}
with the notation of Section~\ref{sec:MIA2}. Analogue relations for other mass insertions can be found accordingly. From this last relation one can also see that the effect disappears in the degenerate case.

The effect of the squark-gluino corrections to the external legs (or the difference between the two definitions of the super-CKM basis), turns out to be numerically important because of the chiral enhancement. While this article was under revision, a phenomenological study of these corrections has been performed in Ref.~\cite{Crivellin:2009ar}, with the conclusion that the bounds on some mass insertions (corresponding to the tree-level definition of the super-CKM basis) are modified considerably. These corrections have obviously no effect on the phenomenological bounds derived for the mass insertions associated to the on-shell definition of the super-CKM basis, and in this scheme the NLO corrections are well defined in the limit of vanishing quark masses. However, it is \emph{not} a claim of the present paper that the on-shell scheme is in any way preferred to the tree-level one. In fact, the tree level scheme might be more convenient if one wishes to relate phenomenological bounds on these type of low energy processes to specific mechanisms of SUSY breaking. This issue has been already discussed in Ref.~\cite{Crivellin:2009ar}. In those cases one can add the flavor-changing self-energies as given in that paper to our NLO results, or do the substitutions of Eqs.~(\ref{rel1}) or (\ref{relDelts}) directly.



\end{document}